\documentclass[manuscript,screen,acmsmall]{acmart}
\AtBeginDocument{%
  \providecommand\BibTeX{{%
    \normalfont B\kern-0.5em{\scshape i\kern-0.25em b}\kern-0.8em\TeX}}}

\setcopyright{acmcopyright}
\copyrightyear{2018}
\acmYear{2018}
\acmDOI{10.1145/1122445.1122456}

\acmConference[CSCW '21]{CSCW}{}{}
\usepackage{url}
\usepackage{paralist}
\usepackage{bm}
\usepackage{booktabs}
\usepackage{multirow}
\usepackage{xcolor}
\PassOptionsToPackage{svgnames,x11names,dvipsnames}{xcolor}
\usepackage[most]{tcolorbox}
\usepackage{graphicx}
\usepackage{enumitem}
\usepackage{wrapfig}
\usepackage{tabularx,array}
\usepackage{bbding}

\usepackage{colortbl}

\definecolor{cback}{HTML}{EDEFF1}
\definecolor{cframe}{HTML}{B9C4CA}
\newtcbox{\inlinebox}[1][]{enhanced,
 box align=base,
 nobeforeafter,
 colback=cback,
 colframe=cframe,
 size=small,
 left=0pt,
 right=0pt,
 boxsep=2pt,
 #1}

\usepackage{algorithm}
\usepackage{algorithmic}

\newcommand{\name}{MTV}

\newcommand{\revise}{\textcolor[rgb]{0, 0, 0}}
\newcommand{\minorrevise}{\textcolor[rgb]{0, 0, 0}}
\newcommand{\cc}[1]{\textcircled{\small{#1}}}

\begin{document}

% \title{MTV: Detecting, Examining, and Labeling Anomalieswith Multivariate Time-Series Visualization}
% \title{MTV: An Integrated Workflow for Detecting, Investigating, and Annotating Anomalies in Multivariate Time Series}
% \title{MTV: Visual Analytics for Collaborative Investigating and Annotating Anomalies in Multivariate Time Series}
\title{MTV: Visual Analytics for Detecting, Investigating, and Annotating Anomalies in Multivariate Time Series}


\author{Dongyu Liu}
\email{dongyu@mit.edu}
\orcid{0000-0002-8915-2785}
\affiliation{%
	\institution{Massachusetts Institute of Technology}
	\streetaddress{77 Massachusetts Ave}
	\city{Cambridge}
	\state{MA}
	\postcode{02139}
}

\author{Sarah Alnegheimish}
\affiliation{\institution{Massachusetts Institute of Technology}}
\email{smish@mit.edu}

\author{Alexandra Zytek}
\affiliation{\institution{Massachusetts Institute of Technology}}
\email{zyteka@mit.edu}

\author{Kalyan Veeramachaneni}
\affiliation{\institution{Massachusetts Institute of Technology}}
\email{kalyanv@mit.edu}

% \affiliation{%


\renewcommand{\shortauthors}{Liu et al.}

\begin{abstract}
    Detecting anomalies in time-varying multivariate data is crucial in various industries for the predictive maintenance of equipment. Numerous machine learning (ML) algorithms have been proposed to support automated anomaly identification. However, a significant amount of human knowledge is still required to interpret, analyze, and calibrate the results of automated analysis.
    This paper investigates current practices used to detect and investigate anomalies in time series data in industrial contexts and identifies corresponding needs.
    Through iterative design and working with nine experts from two industry domains (aerospace and energy), we characterize six design elements required for a successful visualization system that supports effective detection, investigation, and annotation of time series anomalies. 
    We summarize an ideal \minorrevise{human-AI collaboration} workflow that streamlines the process and supports efficient and collaborative analysis.
    We introduce \name~(\underline{M}ultivariate \underline{T}ime Series \underline{V}isualization), a visual analytics system to support such workflow.
    The system incorporates a set of novel visualization and interaction designs to support multi-faceted time series exploration, efficient in-situ anomaly annotation, and insight communication.
    Two user studies, one with 6 spacecraft experts (with routine anomaly analysis tasks) and one with 25 general end-users (without such tasks), are conducted to demonstrate the effectiveness and usefulness of \name.
\end{abstract}




\begin{teaserfigure}
  \includegraphics[width=\textwidth]{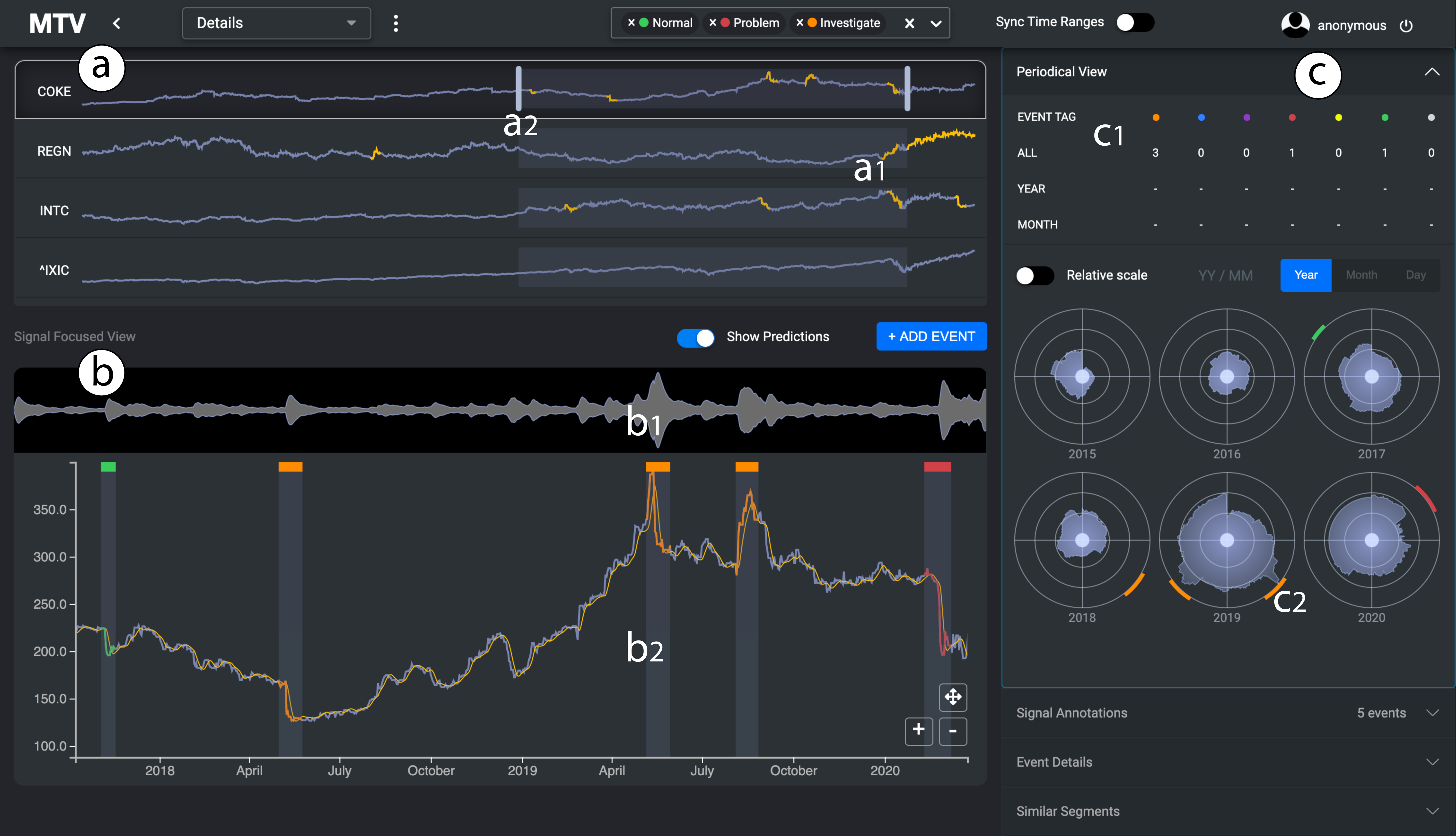}
  \caption{The \name~interface, \revise{displaying an analysis of stock data. The Signal Overview (a) displays multiple signals (in this case, stocks) that share the same horizontal timeline and highlights anomalous events with a warning color (a1). Users pick a signal of interest (i.e., COKE) and brush a segment (a2) to observe its details and interact with the events in the Signal Focused View (b). Events can be color-tagged (e.g., green - normal, orange - investigate, red - problem) and filtered in the top header.
  The predicted errors (b1) can be toggled to explain why a certain event (b2) was identified by the machine learning algorithm. The Side Panel (c) includes four collapsible views --- the Periodical View (c1, c2), Signal Annotations View, Event Details View, and Similar Segments View --- which allow users to investigate and annotate events efficiently and collaboratively.}}
  \label{fig:teaser}
\end{teaserfigure}

\maketitle

\section{Introduction}\label{sec:introduction}
The rapid proliferation of sensors and connected devices has led to a massive, ever-increasing accumulation of temporal observation data (a.k.a. multivariate time series). As more and more industries rely on such data, detecting and analyzing anomalies in time series becomes increasingly important for critical use cases, ranging from cyber-intrusion and fault detection to preventative maintenance and fraud prevention~\cite{ahmad2017unsupervised}. 

A time series anomaly is a time point or period during which a system exhibits unusual behavior. To efficiently detect such unusual behaviors in what are often enormous datasets, and ease the burden on the humans often tasked with such detection, a variety of automated anomaly detection methods using statistical or machine learning techniques have been proposed~\cite{hodge2004survey, chandola2009anomaly, goldstein2016comparative, habeeb2019real}. A typical approach, employed by applied machine learning engineers, is to use an ML technique (possibly a deep learning approach), flag anomalous periods or time points, and send them to  experts/operators, often using simple \texttt{csv} files. 

In practice, however, these methods only solve part of the problem --- detecting unusual behavior. Many real-world systems are highly dynamic, and can only be fully understood by including context and domain expertise outside the scope of the time series analyzed by these methods. Simply put, the presence of unusual behavior alone does not necessarily mean there is a \textit{problem} that needs attention. For example, if a satellite is passing an eclipse, its signals (one of the datasets we consider) might exhibit some unusual behaviors that would not necessarily indicate problems. 
Additionally, system dynamics, like maneuvers, mode switching and others, may manifest in signals as unusual but are not necessarily troublesome. In our collaborations with domain experts and operators, we found that their reluctance to rely on the ML-based tactics above stemmed from their inability to integrate their own knowledge, context and analysis with the ML approach, to share their analyses with one another, to understand the outputs of ML approaches, and ultimately to reduce the number of false positives over time. Thus, we set out to study and design a flexible \minorrevise{human-AI collaboration} workflow to overcome this reluctance.
%to detect and investigate anomalies in multivariate time series. 

Data visualization supports this workflow by integrating humans' valuable expertise and creativity into anomaly investigation via various visual representations of information and human-machine interactions~\cite{ware2019information}.
Recent years have seen increasing efforts to combine visualization and automated anomaly detection for time series data analysis~\cite{xu2019clouddet, xie2018visual}.
However, effectively visualizing multivariate time series is difficult, as it requires determining how to get across complex yet central aspects of the data --- including issues of scalability (what is the right level of detail to show), multi-dimensionality (how many time series should be displayed at once), and interpretability (how best to represent the contextual information that helps determine legitimate anomalies). 

Other questions come up as well:
(1) \textbf{End-user:}
A large number of end-users tasked with \revise{time series anomaly analysis} may have minimal or no technical background.
To help them, we must understand what they are hoping to \underline{\textit{observe}} or \underline{\textit{control}} about the data and the analysis process, and avoid using overly technical jargon or complicated visualizations. An intuitive visual design is required to alleviate the learning burden for users.
(2) \textbf{Annotation:} 
From the user's point of view, it is the interpretation of results that matters. However, existing visual systems~\cite{cao2017voila, xu2019clouddet, xu2018ecglens} for time series anomaly analysis lack the ability to document and share such interpretations, without which it can be difficult to translate results into actionable decisions. We stress that the ability to annotate anomalies, as well as to organize and present these annotations, is a highly important part of any anomaly detection system.
(3) \textbf{Workflow:}
Bringing humans into the analysis loop to make sense of ML results is not a simple or one-step task.
Still more questions arise during this process, including: 
What ML outputs does it make sense to start with? 
How should teams collaborate to annotate anomalies? 
Can the ML system use these annotations to reduce false alarms in the future? 
Little existing work has investigated these questions deeply.

Awareness of these gaps comes through our collaboration with domain experts in the aerospace and energy industries, who spend a considerable amount of time monitoring telemetry data from large industrial devices in order to identify and analyze potentially hazardous events.
This work represents an early step in understanding the support-annotating and sensemaking needs of these experts --- who, like many in industrial environments, work with a large number of large-scale time series, each of which may contain more than tens of thousands of data points. In this context, we lay out research questions and summarize our key contributions as follows. The results of RQ1, RQ2, and RQ3 are all derived from \minorrevise{an iterative user-centered design process} which involves 6 experts from aerospace and 3 experts from energy industry. 
\begin{itemize}
    \item \textbf{RQ1.} \revise{\textit{What design elements are required for a system to support time series anomaly analysis in the above-mentioned context?}}
    \\
    \revise{
    We summarize the current challenges commonly faced by both the aerospace and energy experts (Section~\ref{sec:running_example}). Then we identify the design requirements (Section~\ref{sec:design_requirements}) necessary for a time series analysis system that incorporates both the power of human intelligence and the computation capability of machines to solve existing challenges.}

    % \item \textbf{RQ2.} \textit{What is the idea workflow that end-users follow to perform anomaly analysis of large-scale time series data \revise{with the assistance of ML?}}
    \item \textbf{RQ2.} \textit{\minorrevise{What is an ideal human-AI collaboration workflow for efficient and effective detection, investigation, and annotation of anomalies in industrial-scale time series data?
    }}
    % \item \textbf{RQ2.} \textit{\revise{What is an ideal workflow supporting human-AI collaboration, as well as team collaborative analysis?}}
    \\
    We summarize a streamlined \minorrevise{human-AI collaboration workflow} (Section~\ref{sec:workflow}) that allows for easy, flexible, and efficient anomaly analysis of multiple large-scale time series data. 
    %An expert can enter the workflow at any stage and finish a job in just a few steps. More importantly, a team can complete their tasks in a collaborative manner.
    %\minorrevise{Importantly, members from a team can also complete their tasks in a collaborative manner. 
    \minorrevise{We report the usability challenges involved in such a workflow, and discuss the lessons learned that also apply to other human-in-the-loop data analytics scenarios, in Section~\ref{sec:discussion}.}
    
    \item \textbf{RQ3.} \textit{What is an effective system solution that enables such \minorrevise{human-AI collaboration} workflow while incorporating all the design elements learned from RQ1?}
    \\
    \revise{
    We develop \name, a visual analytics system that follows the aforementioned design requirements and demonstrates our proposed workflow.
    The system includes an end-to-end ML pipeline that detects anomalies without labeled data (Section~\ref{sec:mtv_ml_pipeline}) and learns from human feedback (Section~\ref{sec:mtv_improve_ml}).
    The interface (Section~\ref{sec:mtv_visualization}) incorporates several novel and intuitive visualization and interaction designs for multi-faceted and variously granular time series data exploration, as well as anomaly investigation and in-situ annotation and communication. 
    Specifically, a novel shape-matching algorithm (Section~\ref{sec:mtv_shape_matching}) is introduced to enhance annotation efficiency. 
    We quantitatively evaluate the shape-matching algorithm, as well as the system's ability to learn from human feedback, and report the results in Section~\ref{sec:evaluation_algorithms}. 
    % A new metric measuring the similarities between two signals is also proposed to better present patterns in anomalous occurrences. 
    }
    
    \item \textbf{RQ4.} \textit{\revise{How do domain experts, who are well-versed in routine anomaly analysis tasks, perceive the usefulness and usability of such a system? How about general end-users with less experience?}}
    \\
    \revise{We evaluate the usefulness of MTV through several case studies, performed by 6 domain experts from the aerospace industry and using spacecraft telemetry data  (Section~\ref{sec:evaluation_experts}). We also run a usability study using stock data in which 25 general end-users reveal the potential broad benefits of such a system (Section~\ref{sec:evaluation_general}). }
\end{itemize}

% \input{texfiles/12.problem.tex}
%!TEX root = ../main.tex
% !TeX spellcheck = en_US

\section{Related Work}\label{sec:relatedwork}

In this section, we summarize the techniques that are most relevant to our work. We have divided these into four types: automated anomaly detection, visualization for time series anomaly detection, annotating with interactive visualizations, and sensemaking and collaborative visualization.

\subsection{Automated Anomaly Detection}
Over the decades, the rich variety of anomaly types, data types and application scenarios has led to the development of numerous anomaly detection approaches. Several surveys have summarized these techniques~\cite{hodge2004survey, chandola2009anomaly, goldstein2016comparative, habeeb2019real}.

The general goal of anomaly detection is to find unexpected patterns in data.
The simplest approaches are "out-of-limits" methods, which are applied to raw values and flag locations where predefined thresholds are surpassed.
\revise{
However, such methods are not suited for detecting \textit{contextual anomalies} that do not fall within low-density values in global but are anomalous with respect to local values.
To overcome this, advanced approaches have since been developed based on statistics~\cite{zheng2016self, pena2013anomaly, pena2013anomaly}, clustering~\cite{angiulli2002fast, breunig2000lof}, and machine learning~\cite{ahmad2017unsupervised, an2015variational} or deep learning~\cite{zhou2019beatgan, hundman2018detecting, geiger2020tadgan}.}

In our work, we build an end-to-end ML pipeline which integrates three powerful algorithms including Arima, LSTM and TadGAN to handle datasets of different characteristics (Section~\ref{sec:mtv_ml_pipeline}).
Our main goal is to engage humans with the analysis loop, enabling them to make sense of the ML results \revise{(especially the identified contextual anomalies)}, and to support collaborative sensemaking between human experts.
%We focus on large-scale multivariate time series data, an area that is underexplored.

\subsection{Visualization for Time Series Anomaly Detection}

There are numerous techniques for visualizing and structuring time series~\cite{aigner2011visualization}. The key difference lies in how the timeline is encoded. Time is linear, but contains an inherent hierarchical structure of granularities, such as hours, days, weeks, and months. Depending on the \minorrevise{analytical tasks} at hand, different design techniques may be employed. 

A standard method for visualizing a time series is mapping time to the horizontal x-axis and time-dependent variables to the vertical y-axis~\cite{aigner2011visualization}.
Line and area charts are the most common ways to represent time series.
When our focus is on observing cyclic or periodical patterns, a spiral-shaped time axis ~\cite{weber2001visualizing} is a useful time-encoding scheme.
If we put more emphasis on individual dates, a calendar layout is more suitable, and can depict daily, monthly, or yearly value changes~\cite{van1999cluster}.

To encode multivariate time series, techniques including superposed line graphs, braided graphs, small multiples, and horizon graphs can be used~\cite{javed2010graphical}. 
Recently, glyph-based designs have also been explored due to their expressiveness and effective use of screen space~\cite{fuchs2013evaluation, cao2015targetvue}.
\revise{Another typical approach involves using multiple coordinated views, with each view displaying particular aspects of time in order to support a coordinated analysis~\cite{aigner2011visualization}.}
Each of the techniques introduced above has its own merits and is well-suited for certain \minorrevise{analytical tasks}. However, it is difficult to use any one of them for anomaly analysis directly.

Several comprehensive visual analytics systems have been developed for time series anomaly investigation. These systems have been applied to various scenarios, such as air quality monitoring~\cite{qu2007visual}, traffic volume monitoring~\cite{cao2017voila}, electronic healthcare records analysis~\cite{xu2018ecglens}, and cloud computing system performance analysis~\cite{xu2019clouddet}. None of them support anomaly annotation and collaborative analysis.

In this work, we propose a set of hybrid visualization and interaction designs tailored for time series anomaly analysis \revise{with a particular focus on collaborative annotations. All these designs together compose our \name~ system, which supports a streamlined and effective \minorrevise{human-AI collaboration workflow} (Section~\ref{sec:workflow}).} 

%Moreover, these systems have relatively high learning curve and require end users to be technically knowledgable. In this work, we address all of these concerns.  

\subsection{Annotating with Interactive Visualizations}
Annotating (labeling and commenting on) data is frequently supported in computer-supported cooperative work and human-computer interaction (CSCW/HCI)~\cite{bernard2017comparing, bernard2018vial, bardram2018collaborative, miceli2020between}. Annotation tasks are widely observed in many application scenarios, as humans leverage their high-level intelligence and domain expertise to add meaningful context to data. Through visual interfaces, humans can annotate interesting entities in textual documents~\cite{xia2014annotatedtimetree, heimerl2012visual}, images~\cite{bernard2017visual}, and videos~\cite{hoferlin2012inter}.
Unlike text and image data that \revise{are} often easily understood by humans, time series data \revise{are} much harder to annotate, and studies regarding time series data annotation are scarce~\cite{sarkar2016visual, alsallakh2014visual}. 
%To the best of our knowledge, no existing work focuses on annotating anomalies in multivariate time series data.
To the best of our knowledge, no existing work focuses on large-scale multivariate time series data annotation through the lens of human-machine collaboration and collaborative analysis among human experts.

Combining machine learning methods and visualization techniques can significantly improve annotation efficacy~\cite{heimerl2012visual, hoferlin2012inter, settles2011closing}.
Unsupervised techniques are often employed to ease the burden of annotation tasks by recommending important instances to annotate.
One typical approach is to use dimensionality reduction techniques to plot instances in the 2D plane, along with instance selection interactions, to allow for easy annotation~\cite{bernard2017comparing}.
Our goal is to investigate time series anomalies and create meaningful annotations for them, which requires having contextual temporal information for anomalous segments. 
Thus, we propose the use of multiple coordinated views, with each view showing different contextual temporal information, \revise{to support in-situ annotation and communication. We further present several novel techniques, including a Multi-Aggregation Viewer and a Similar Segments View powered by shape-matching algorithms, to facilitate anomaly annotation and boost efficiency}.
%Based on this, we specifically propose an investigation workflow to summarize the whole process of time series anomaly investigation.

\subsection{Sensemaking and Collaborative Visualization}

Collaborative visualization is defined as ``\textit{the shared use of computer-supported, (interactive) visual representations of data by more than one person with the common goal of contribution to joint information processing activities}'' by Isenberg et
al.~\cite{isenberg2011collaborative}.
% Prior research in CSCW/HCI has demonstrated the importance of collaborative visualization for sensemaking across various data types, such as tweet data~\cite{torkildson2014analysis}, mobile data~\cite{ludwig2015collaborative}, sound data~\cite{dema_collaborative_2017}, crime data~\cite{goyal2016effects} \minorrevise{and web search results~\cite{morris2013collaborative}.}
%cosearch
\minorrevise{Prior research in CSCW/HCI has demonstrated the importance of collaborative visualization for information-seeking and decision-making~\cite{morris2007searchtogether,amershi2008cosearch,isenberg2009collaborative, shah2012collaborative, morris2013collaborative, hong2018collaborative, liu2018consensus, hong2019design}, and for sensemaking of a variety types of application data 
such as tweet data~\cite{torkildson2014analysis}, mobile data~\cite{ludwig2015collaborative}, sound data~\cite{dema_collaborative_2017} and crime data~\cite{goyal2016effects}}

Sensemaking in collaborative visual analytics is challenging. Experts must have iterative discussions in order to form and verify hypotheses, draw conclusions, and publish findings~\cite{mahyar2014supporting}.
Members of a team must also maintain \textit{awareness} of each other's work in order to progress~\cite{gava20123c}. 
(In our scenario, the ``work'' refers to annotations on anomalous events.)
\minorrevise{Many existing tools enable team members to record questions and insights --- in the form of text, audio, or visual diagrams --- to facilitate organizing and sharing their results~\cite{brennan2006toward, viegas2006communication, heer2008creation, isenberg2011co, willett2011commentspace, hajizadeh2013supporting, mahyar2014supporting}. }
%Other work has explored supporting intra-team awareness to facilitate communication and coordination .

\minorrevise{Compared with synchronous collaboration (through shared workplaces or real-time networked displays)~\cite{isenberg2007interactive,isenberg2009collaborative, isenberg2011co, mahyar2014supporting}}, remote asynchronous collaboration~\cite{heer2007voyagers, heer2008design, zhao2017supporting} remains relatively unexplored, particularly in the context of time series data.
\minorrevise{
Compared with text or images, time series are intrinsically difficult for humans to interpret without sufficient contextual information. 
This makes it even harder for annotators to create accurate externalizations (external representations of a person's internal thoughts) in order to communicate.
How to encode and display these externalizations, and how to maintain awareness between team members in a remote asynchronous setting, are still open research questions.
}
%In this work, we investigate how remote asynchronous collaboration can facilitate time series anomaly analysis by teams.
In this work, we summarize a streamlined \minorrevise{human-AI collaboration workflow} that supports efficient collaborative anomaly annotations. More importantly, we propose a set of novel visualizations and interactions to support in-situ annotation and communication, as well as to enhance team members' awareness of each other's findings.
%\input{texfiles/goals_requirements.tex}

%!TEX root = ../main.tex
% !TeX spellcheck = en_US

\section{Method}\label{sec:background}

\subsection{Iterative Design}
We followed \minorrevise{an iterative user-centered design process} to develop \name.
We collaborated with industrial domain experts, gathering design requirements and collecting feedback from them.
We began with proof-of-concept mockup designs made using Figma (an online collaborative interface design tool). We then moved on to an interactive system prototype, a high-fidelity prototype, and eventually to a deployable system.    

\subsubsection{\textbf{Participants:}}
\label{sec:design_experts}
The entire design process involved 9 participants in two groups. 

The primary group consisted of 6 domain experts from a world-leading communication satellite company: one spacecraft program manager (P1) and five senior satellite engineers (P2-P6).
%All are responsible for providing anomaly support to the Spacecraft Operation Center in their company. 
The team analyzes spacecraft telemetry data for signs of hazards that may result in system failure.
Each expert has between 5 and 17 years of spacecraft telemetry data analysis experience, and between 0 and 3 years of machine learning experience. This group participated in our regular design meetings, which occurred once or twice per month for three years, and provided feedback on prototypes. 

The external group included \revise{3 senior engineers (E1-E3) from an energy company with expertise in analyzing time series from wind turbines and energy pipelines. They took surveys about the system design and usability, and tested \name~ with their own data}. This group provided assurance that the design requirements and features we developed with the primary group are \revise{generalizable in certain contexts (defined in Fig.~\ref{fig:position})}, as well as applicable to similar fields.

\subsubsection{\textbf{Design stages:}}
The design process occurred in three stages.

\revise{We began the first stage by conducting interviews with the experts from the primary group (P1-P6) to understand in detail the methods they use to analyze spacecraft telemetry data for anomalies as part of their routine workload, along with the associated challenges. We detail this process in Section~\ref{sec:running_example}.
Knowing all these challenges, we had additional conversions with E1-E3 and confirmed that the same challenges also exist in their domain. 
}

In this first stage, we gathered initial requirements and designed mockups using Figma by choosing different combinations of visualization techniques in accordance with design principles. We refined the requirements iteratively, and finally derived a set of requirements to guide our development of the early version of \name.   

In the second stage, we focused on implementing \name~ and iteratively improving it by engaging domain experts from the primary group. We then ran our high-fidelity prototype on public data. This process involved many informal discussions and interviews. From time to time, we also presented the prototype to E1-E3 and collected feedback from them. The output of this stage was a high-fidelity software prototype.      

Lastly, we deployed our prototype using real spacecraft telemetry data in a production environment, and conducted four case studies with P1-P6 through online meetings with screen sharing. Based on the feedback collected in each case study, we continued to refine our system. The experiment details and results are reported in Section \ref{evaluation:experts}.

\subsection{\revise{How Satellite Experts Detect and Investigate Anomalies in Spacecraft Telemetry?}}
\label{sec:running_example}

\revise{
One major objective of the satellite company's operative team is to detect unexpected behaviors (i.e., anomalies) in tens of thousands of signals. Through our three-year collaboration with the satellite company, we formed an understanding of the scope of this task and what it entails.
The team works with multiple spacecrafts. Each spacecraft telemetry database contains around 37,000 signals spanning 9 different subsystems from one spacecraft. Each signal is a univariate time series collected at the microsecond level, and has been tracked for over 10 years.}

\revise{
The team's conventional approach to anomaly detection is based around setting and adjusting thresholds in order to flag anomalous intervals.
The team then manually reviews suspicious intervals, often using simple \texttt{csv} files, and examines individual signals in a third-party platform such as \texttt{MatLAB}.
For each anomaly detected, the team digs into the corresponding signal to track the root cause of the alarm. During this investigation process, some relevant signals (usually from the same subsystem) will be examined. An average of 20 alarms is reported every day, most of which are false alarms and can be resolved within a few hours.
%For some that are identified as true alarms but non-urgent, the experts create a daily report (a text document) and set up in-person team meeting every several days to solve issues one by one.
For some that are identified as true alarms but non-urgent, the experts gather further information over some time window to help explain the root cause and the way forward.
}

\minorrevise{Over the course of this process, the team faced many challenges:}

\begin{enumerate}[label=\textbf{C\arabic*}]

\item
The team's current method is highly expensive, and forces them to restrict their focus to a subset of a few hundred signals chosen based on domain knowledge.

\item
The team has wanted to use ML models to identify contextual anomalies --- anomalies that do not exceed a normal range, but are unusual compared to local values. But they have limited machine learning experience and have not been able to find a way to do this.

\item
The team found that ML models often flag unusual patterns, even when these patterns do not necessarily indicate a problem. For example, an eclipse might cause patterns that are then flagged even though they are not troublesome. The team is eager for their models not to mark these patterns, but struggles to integrate this feedback.

\item
To track the root cause of an anomalous event, the team often needs to observe several relevant signals. As such, they have to frequently shift between different tools such as \texttt{Excel} and \texttt{MatLab}. This process becomes particularly inefficient when the number of signals that need to be analyzed together increases.

\item
Frequently, the team needs informal discussion between team members. However, it is arduous for the team to document what they found and communicate their insights effectively, which is the same challenge they face during the in-person meeting.
%\textbf{C6} Once a team member leaves, his/her knowledge or expertise will not stay. The team desires to have a 
\end{enumerate}

\subsection{Design Requirements}
\label{sec:design_requirements}

\revise{Motivated by the aforementioned challenges, we came up with the following design requirements, which guided the development of \name.}

\begin{enumerate}[label=\textbf{R\arabic*}]
    \item \textbf{Facilitate efficient and intelligent anomaly identification (C1, C3).}
    \\
    \revise{The scale of time series data is often overwhelming in terms of either the number of signals or the number of data points per signal.}
    % To maintain a complex system such as spacecraft, over thousands of time-varying measurements  (e.g., temperature, radiation, and power) are being collected in a second/minute granularity.
    The system should support fast and robust automated anomaly detection.
    In addition, the system is expected to learn from human annotations (e.g., tags and comments) of existing anomalous events in order to avoid generating too many false alarms or missing true anomalies.
    %In addition, the system is expected to learn from human's interactions (e.g. tag/comment/edit/create/delete) with existing anomalous events to know how to intelligently point experts to the most important segment of time series, prioritize them, and provide information about why they are important.
    
    \item \textbf{Provide the flexibility to interact with \revise{ML results (C2)}.}
    \\
    \revise{ML can exhaust every meaningful hyperparameter combination suggested by experts.
    Different settings may lead to different results that make sense in certain analytical scenarios.
    For example, the satellite experts may want to analyze ML results over different time aggregation levels, such as "6 minutes" and "1 hour" (Section~\ref{sec:system_pipeline}).
    The system should save possibly meaningful ML results and allow experts to interact with these results in real time to choose what they want to investigate.
    }
	
	\item \textbf{Offer a visual interface to \revise{display and interact with ML-flagged events (C3, C4).}}
	\\
	The visual interface should display when and where the anomalies occurred and \revise{point experts to the most important time segments, prioritized by severity.}
	Create, read (aka retrieve), update, and delete (CRUD) functions should be supported when experts interact with an event. For example, when an expert \revise{finds} a segment important but it is not flagged by the machine, s/he should be allowed to easily create an event on this segment.
	
% 	More importantly, collaborative exploration must be supported given heavy workload brought from the extremely large size of real data.
% 	To this end, interaction history is suggested to be recorded and traceable through the interface so that all experts can know the sufficient details of an event and take well-informed action.
% 	In addition, the system should further allow experts to document their thoughts, make annotations on anomalous events, and explore annotations from other experts. 
	
	\item \textbf{\revise{Allow efficient in-situ annotation and communication (C5).}}
	\\
    \revise{The system should allow experts to document their thoughts, as well as to assign tags to an event. We refer to this behavior as event annotation.
	In-situ annotation~\cite{bardram2018collaborative} --- in our scenario, annotating directly on the time segment of the event under investigation --- should be supported.
	The system should also allow in-situ communication of insights (i.e., tags and textual comments) and enhance team members' awareness of each other's findings. This helps experts to make well-informed annotations, and to verify each other's decisions with their complementary knowledge. 
	Techniques should be used to assist with efficient annotation, as the large-scale nature of the data may lead to too many anomalies being identified.
	}
	
	\item \textbf{Enable multi-faceted and multi-granular visual explorations of time series (C4).}
	\\
	To interpret and annotate anomalies, the system should provide experts with contextual information about \revise{the anomalies}, allowing them to confirm or refute the system's conclusions.
	Therefore, the ability to show time series patterns from different perspectives (trend, periodicity, etc.) and in different granularities (second, minute, hour, etc) is required.
	
    \item \textbf{\revise{Afford a streamlined workflow for efficient collaboration (C1-C5).}}
    \\
    \revise{
    Experts are expected to follow a linear workflow to perform anomaly analysis.
    First, the workflow should meet all the previous requirements (\textbf{R1-R5}).
    Second, experts are allowed to enter the workflow at any stage in order to do their job without starting from scratch.
    For example, expert A may specify what ML results to investigate, while experts B and C perform detailed analysis and make annotations collaboratively and expert D engages in discussion around an anomaly; expert A may then come back to check the final annotation results.
    In short, the pathway to any individual task should be directly accessible. 
    }
    %Mind stress to learn the tool for domain experts like Jean -> go linearly -> embed their workflow.
    %[respond to their job not just for research]
    %[get into the real end-users]
   % [the first-hand people]
    
\end{enumerate}

\begin{table}[!htp]
\def\arraystretch{1.35}
\begin{tabular}{@{}p{2.5cm} p{11.5cm}@{}}
\multirow{2}{*}{\includegraphics[width=2cm]{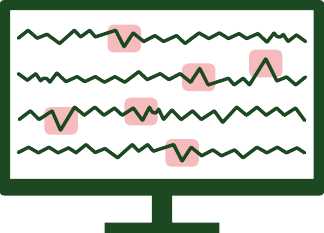}}
& \textbf{T1. Extract anomalies from massive time series (R1). \inlinebox{ML} }\label{t:extract_ml}
\\ &
The first step is to use ML to detect anomalous events within our dataset, which consists of \revise{multiple (more than 100) large-scale time series (often more than 10K data points/time steps) without labeled data}. An end-to-end \revise{unsupervised} ML pipeline is built for this purpose. 
The pipeline exposes analytical-task-relevant hyperparameters to experts, such as the time interval for aggregating raw signals and strategies to impute missing values.
The ML results of every meaningful hyperparameter setting are saved.
\\
\multirow{2}{*}{\includegraphics[width=2cm]{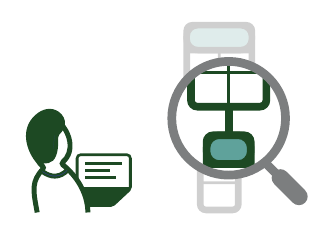}}
& \textbf{T2. Select signals and ML results to investigate (R2). \inlinebox{Human}}\label{t:select_ml}
\\ &
Depending on need, an expert selects a subset of signals (\revise{a few to dozens}) and ML pipeline outputs to investigate. For example, s/he may want to look at anomalies identified in all temperature signals ($\sim$20 of them) aggregated at an one-hour level. 
% This selection happens on the "landing page" (Sec~\ref{sec:landing_page}).
\\
\multirow{2}{*}{\includegraphics[width=2cm]{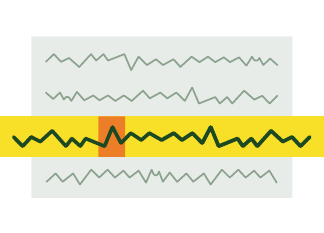}}
& \textbf{T3. \revise{Scan through these signals and pick one of interest} (R3, R5). \inlinebox{Human}}\label{t:scan_signals}
\\ &
\revise{After selecting the signals, the expert may want to observe an overview of them in which multiple time series are plotted and anomalous events are highlighted. This allows s/he to quickly scan through the dynamics of multiple signals and then decide which to pick for further investigation.}
\\
\multirow{2}{*}{\includegraphics[width=2cm]{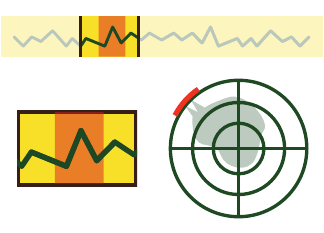}}
& \textbf{T4. Explore one signal and \revise{its anomalies} in depth (R4, R5). \inlinebox{Human}}\label{t:explore_one}
\\ &
\revise{Once the expert finds the signal and anomalies of interest, s/he may explore contextual information about the anomalies. This information includes but is not limited to: (1) the surrounding time series values of the anomaly at different aggregation levels; (2) the periodical (daily/monthly/yearly) patterns of the time series; (3) any annotations from other team members; and (4) an anomaly's total interaction/annotation history.}
\\
\multirow{2}{*}{\includegraphics[width=2cm]{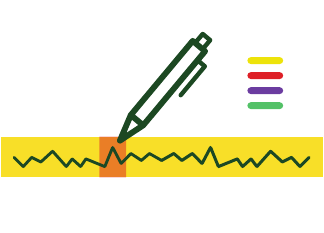}}
& \textbf{T5. \revise{Annotate the anomalies from this time series (R3, R4).} \inlinebox{ML+Human}}\label{t:annotate_one}
\\ &
\revise{
After gaining a sufficient understanding of one anomalous event, the expert may interact with (CRUD) or annotate (tag and comment) the event.
If this event has already been annotated by other team members, the expert can still add more comments and change the tag if necessary.
Given the large-scale nature of the data, it is challenging for experts to annotate all events. Automatic techniques such as shape-matching are expected to boost the efficiency of annotation. 
}
\\
\multirow{2}{*}{\includegraphics[width=2cm]{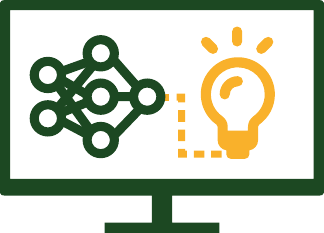}}
& \textbf{\revise{T6. Enhance ML pipeline using annotations (R1).} \inlinebox{ML+Human}}\label{t:enhance_ml}
\\ &
\revise{All annotations and CRUD interactions are stored in the database. These are leveraged to improve the performance of the ML pipeline, forming a closed loop.
In other words, we incorporate experts' knowledge into the system so that it can learn from event patterns, mitigating false alarms in favor of surfacing true anomalies in the future.
}

\end{tabular}
\caption{ 
\revise{The ideal \minorrevise{human-AI collaboration workflow} that streamlines tasks. It starts with ML extracting anomalies from massive time series, continues with users giving high-level input to machines about what to observe, proceeds with an overview-first and details-on-demand exploration and a fast in-situ annotation strategy, and ends with enhancing the ML pipeline through annotations, forming a closed loop.}
% so that experts can enter the workflow at any stage to do their job without starting from scratch.}
}
\label{tb:workflow}
\end{table}

\subsection{Workflow}
\label{sec:workflow}

\begin{figure}[!htbp]
	\centering
	\includegraphics[width=0.9\linewidth]{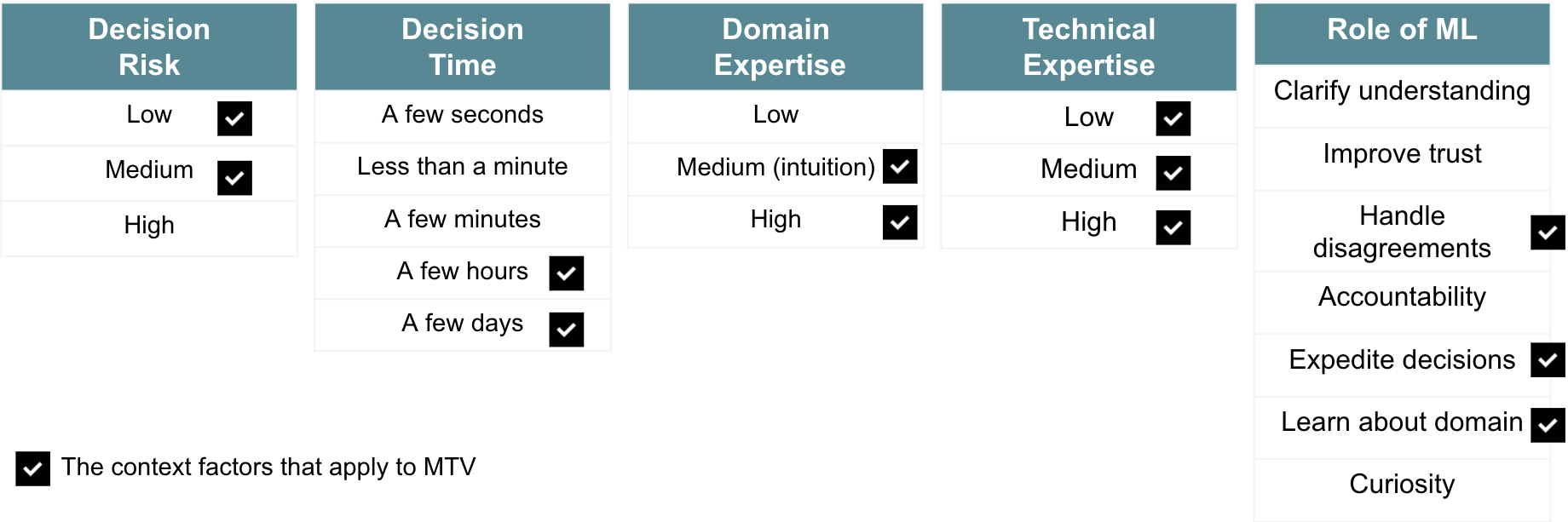}
	\caption{\revise{The five domain context factors we used to determine the scenarios in which our proposed workflow and system are a good fit for time series anomaly analysis.
	\textit{Decision} refers to the process through which a team judges an ML-identified anomaly and provides a corresponding explanation.
	\name~is designed to support asynchronous collaborative analysis, and is thereby suited for decisions of that are of low or medium risk and take at least a few hours. The system is friendly to non-technical users but requires a certain amount of domain expertise so that time series can be interpreted.}}
	\label{fig:position}
	\vspace{-1ex}
\end{figure}

\revise{
Through iterations, we identify an ideal \minorrevise{human-AI collaboration workflow} (\textbf{R6}) that streamlines the process of time series anomaly analysis.
Experts can enter the workflow at any stage and finish their tasks in an efficient and collaborative manner.
Table~\ref{tb:workflow} summarizes all the relevant steps in detail. The workflow starts with ML extracting anomalies from massive time series, continues with users giving high-level input to machines about what to observe, \revise{proceeds with an overview-first and details-on-demand exploration (a.k.a. Shneiderman's mantra ~\cite{shneiderman2003eyes}) and a fast in-situ annotation strategy}, and ends with enhancing the ML pipeline through annotations, forming a closed loop.
}

It is worth noting that this workflow is not meant to fit every single analytical scenario.
To clarify its scope, we consider five domain context factors (shown in Fig.~\ref{fig:position}) inspired by the work of Zytek et. al.~\cite{zytek2021sibyl}. 
To instantiate this workflow, we developed MTV, an interactive visual analytics system for time series anomaly analysis at an industrial scale. The system supports the foregoing \minorrevise{human-AI collaboration workflow} and integrates all the features to meet the design requirements.

%!TEX root = ../main.tex
% !TeX spellcheck = en_US

\section{MTV}\label{sec:system}

In this section, we first describe our approach of identifying anomalies through an end-to-end machine learning pipeline, followed by the introduction of improving ML with annotations.
Next, we detail how we use shape-matching techniques to enhance annotation efficiency.
Finally, we describe the visual design of our system.

\subsection{Identify Anomalies with the End-to-end ML Pipeline }
\label{sec:mtv_ml_pipeline}
\label{sec:system_pipeline}

\begin{figure}[!htbp]
	\centering
	\includegraphics[width=1\linewidth]{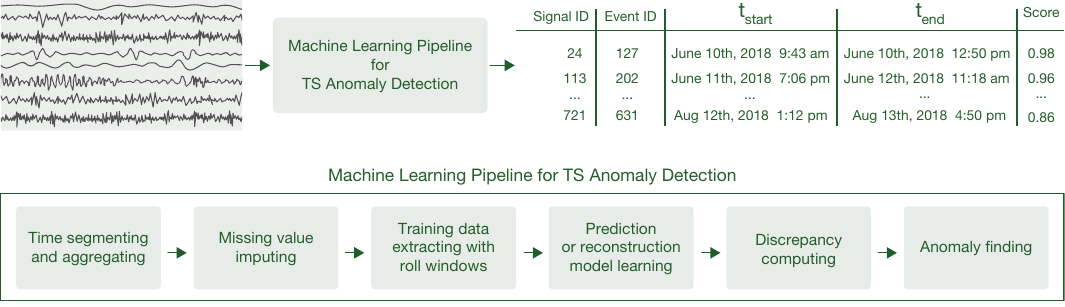}
	\caption{The end-to-end ML pipeline for multivariate time series anomaly detection. (Top) The pipeline takes multiple signals as inputs and generates a list of intervals of timestamps ($t_s$, $t_e$) suspected to be anomalous. (Bottom) The pipeline is composed of six primitives, each serving one particular function.}
	\label{fig:ml_pipeline}
	\vspace{-1ex}
\end{figure}

% Detecting anomalies within time series data does not simply mean training a prediction/reconstruction model. Numerous pre-processing and post-processing steps should be applied as well.
Fig.~\ref{fig:ml_pipeline} describes an entire ML pipeline from start to finish. 
At a high level, the pipeline takes multiple signals as inputs and generates a list of intervals of timestamps ($t_s$, $t_e$) which are suspected to be anomalous (\textbf{T1}).
%A primitive is a reusable, self-contained, software component for machine learning~\cite{smith2020machine}.
The pipeline contains six fundamental primitives~\cite{smith2020machine}, each serving a particular function, and each with various hyperparameters, the settings of which influence the ML outputs\footnote{The url (\url{https://bit.ly/mtv_ml_pipelines}) links to our github repo (\textbf{anonymized for review purpose}) where we detail how the hyperparameters of each primitive of the LSTM pipeline are set. More pipelines can be found under folder \texttt{XXXX/pipelines/verified/}.}. Different settings will result in different ML outputs that are meaningful depending on different analytical demands. Because domain experts generally do not have much ML expertise, we have designed our system around what they hope to observe (\textbf{T2}), and provided them with control through the Landing Page (Section~\ref{sec:landing_page}). 

\textbf{Data pre-processing.}
The first two primitives transform the data into a clean format $\mathbf{x} = [x^1, x^2, \dots, x^T]$, \revise{where $T$ denotes the total number of time steps} and the intervals between  $x^{t-1}$ and $x^{t}$ are equal.
\revise{To fill any missing values, various predefined imputation strategies are supported, with the mean value used by default.
For spacecraft telemetry data, we use intervals from 6 minutes to 6 hours and a zero-order hold strategy\footnote{Zero-order hold: \url{https://en.wikipedia.org/wiki/Zero-order_hold}} to fill missing values according to the domain experts' request. 
}
The third primitive prepares a collection of training samples using a sliding window sequence approach, \revise{and produces $N$ subsequences $X = \{x_{i}^{1\dots t}\}_{i=1}^N$, where $N = (T - t)/ s$ , where $t$ and $s$ represent window size and step size respectively.}

% For a window of size $t$ and step size $s$, we generate $N$ subsequences $X = \{x_{i}^{1\dots t}\}_{i=1}^N$, where $N = (T - t)/ s$.

\textbf{Modeling.}
The fourth primitive learns a prediction/reconstruction model to generate a predicted/reconstructed time series $\hat{\mathbf{x}}$. 
We identify the need to use different algorithms in different scenarios while collaborating with domain experts.
~\name~ has integrated three popular models:
the GAN model (TadGAN~\cite{geiger2020tadgan}) performs well at handling complex signal data with myriad fluctuations; LSTM~\cite{hundman2018detecting} is a well-established sequence learning method suitable for many general scenarios; and Arima~\cite{pena2013anomaly} is a statistical model proven to work excellently when time series have remarkable trend and periodical patterns.
Bearing this high-level knowledge in mind, experts can choose those ML outputs that apply the most suitable model for further investigation (Section~\ref{sec:landing_page}).

\textbf{Finding anomalies.}
Next, the fifth primitive computes the discrepancies between $\mathbf{x}$ and $\hat{\mathbf{x}}$ to locate potential anomalies, under the logic that higher discrepancies suggest a higher chance that the segment is anomalous. In other words, we generate $\mathbf{E} = [e^1, e^2, \dots, e^T]$ to measure the difference at every time step. Then we apply an exponentially weighted moving average (EWMA) on it, obtaining the smoothed error as below: 
\begin{equation}
\mathbf{E_s} = [e^{1}_s, e^{2}_s, \dots, e^{T}_s]
\label{eq:errors}
\end{equation}
The last primitive takes this error sequence as an input and computes a threshold. Any values regarding the
smoothed errors above the threshold are considered to be anomalies.
The threshold is selected from the set: 
$\bm{\theta} = \mu(\bm{e}_s) + k\sigma(\bm{e}_s)$
where $\mu$ and $\sigma$ denote the mean and standard deviation respectively. 
%%%% easy language %%%%
Eventually, $\theta$ is determined by finding a threshold that would bring about the maximum percent decrease in the mean and standard deviation of $\bm{e}_s$ if all error values above the threshold were eliminated~\cite{hundman2018detecting}. Now each anomalous sequence $\bm{e}_{seq}$ (continuous sequences of $e^{i}_s$ with the value above $\theta$) can be assigned a severity score $s$:
\begin{equation}
s = \frac{max(\bm{e}_{seq}) - \theta}{\mu(\bm{e}_s) + \sigma(\bm{e}_s)}
\label{eq:score}
\end{equation}
Now we can represent each anomalous sequence $\bm{e}_{seq}$ in the format of ($t_s, t_e$, $s$), where $t_s$ is the starting time of this sequence, $t_e$ is the ending time, and $s$ is the assigned severity score.

\subsection{\revise{Improve Machine Learning with Annotations}}
\label{sec:mtv_improve_ml}

\begin{figure}[!htbp]
	\centering
	\includegraphics[width=0.65\linewidth]{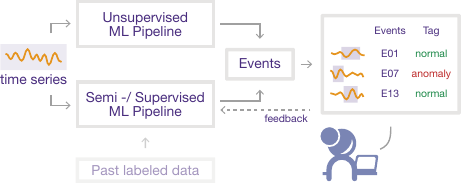}
	\caption{An illustration of how \name~ learns from expert  feedback and improves ML performance over time. }
	\label{fig:improve_ml}
	\vspace{-1ex}
\end{figure}

\revise{
An unsupervised ML pipeline detects anomalies that generally show up as unexpected temporal patterns within certain time periods.
As shown in Fig.~\ref{fig:improve_ml}, in the early phase when there is a lack of annotated data, an unsupervised ML pipeline is used to locate anomalous events.
These anomalies are presented to experts for annotation.
The annotated events are then provided to the semi-/supervised LSTM pipeline, which can be optionally pre-trained with past labeled data. This pipeline keeps learning from feedback and improving on performance over time (\textbf{T6}).
}

\revise{
We base the refreshment process of the semi-/supervised pipeline on application-specific batch processing of annotations. We justify our design based on the high variability between domains in terms of the frequency of anomalies. For example, based on the satellite company's particular needs, we decided that a weekly update is sufficient.
}

\subsection{Enhance Annotating Efficiency with Shape-Matching}
\label{sec:shape_matching}
\label{sec:mtv_shape_matching}
Due to the large-scale nature of the data, experts are only able to annotate a limited number of events. 
We use the idea of ``shape-matching'' to enhance annotating efficiency (\textbf{T5}), as well as to improve the performance of the ML pipeline (\textbf{T6}).
We introduce a novel shape-matching algorithm for efficient shape search within time series data. 

The algorithm is performed in a signal-based manner and outputs a set of candidate shapes in a signal $\mathbf{x}$ that are similar to a particular sub-sequence $\mathbf{s}$, under the constraint that the returned segments will not overlap with each other or any existing events. \revise{Such constraint is important without which experts will feel confused by overlapped events.
We adopt a sliding window approach to generate subsequences of $\mathbf{x}$, then compare them to $\mathbf{s}$ through a similarity measure $f(\cdot, \cdot)$.
We chose $f$ to be the total cost of finding the optimal mapping between two sequences, using either Euclidean or Dynamic Time Warping (DTW~\cite{berndt1994using}) as the similarity measurement. The Euclidean method tends to find exactly matched shapes, while DTW can tolerate a certain level of shifting. 
}

The full algorithmic process is described in the following pseudo-code (algorithm~\ref{alg:shape-matching}).
At each checkpoint $t_c$, we attempt to include the current most similar shape $\mathbf{c}$ in the candidate subset.
Prior to that, we check whether $\mathbf{c}$ overlaps with a preexisting candidate shape; if so, we keep the most similar of the two.
This procedure will return a set of non-overlapping candidate shapes $C$ that are most similar to $\mathbf{s}$.

\begin{algorithm}[!htb]
   \caption{Shape Matching}
   \label{alg:shape-matching}
    \begin{algorithmic}[1]
       \STATE {\bfseries Input:} Signal $\mathbf{x} = [x^1, x^2, \dots, x^T]$
       \STATE Sub-Sequence $\mathbf{s} = [s^{k+1}, s^{k+2}, \dots, s^{k+t}] $ where $ 1 \leq k < (T-t)$.
       \STATE {\bfseries Output:} Candidate sub-sequences $C$.
       \STATE Initialize $\mathbf{c} = \texttt{None}$ \hfill \texttt{// to hold candidate shape}
       \STATE Initialize $t_c = t$.
       \FOR{$i=1$ {\bfseries to} $T-t$}
       \STATE $\Tilde{\mathbf{x}} = [x^i, x^{i+1}, \dots, x^{i+t}]$ \hfill \texttt{// subsequence of $\mathbf{x}$}
       \IF{$f(\Tilde{\mathbf{x}}, \mathbf{s}) < f(\mathbf{c}, \mathbf{s})$}
       \STATE $\mathbf{c} \gets \Tilde{\mathbf{x}}$ \hfill \texttt{// update $\mathbf{c}$ to hold a more similar shape}
       \ENDIF
       \IF{$i > t_c$} 
       \IF{$C \cap \{\mathbf{c}\} \neq \varnothing$}
       \STATE $\mathbf{v} \gets C \cap \{\mathbf{c}\}$ \hfill \texttt{// overlapping sequence}
       \STATE $\mathbf{c} \gets \texttt{argmin} \: [f(\mathbf{c}, \mathbf{s}), f(\mathbf{v}, \mathbf{s})]$ \hfill \texttt{// keep most similar}
       \ENDIF
       \STATE $C = C \cup \{c\}$ \hfill\texttt{// add candidate shape}
       \STATE $\mathbf{c} = \texttt{None}$ \hfill \texttt{// reset}
       \STATE $t_c = t_c + t$
       \ENDIF
       \ENDFOR
       \STATE \textbf{return} $C$
    \end{algorithmic}
\end{algorithm}

We argue for three main uses for shape-matching (described below), all of which greatly enhance annotating efficiency and make anomaly investigation and annotations feasible for large-scale data.

\textbf{Annotation sharing.} 
Once an expert confirms that an annotated sequence is indeed an anomaly, s/he will be inclined to search for other segments that are similar to the confirmed event, but have not yet been flagged.
These similar segments are likely to receive the same annotations as the confirmed one.  
Fig.~\ref{fig:6657}\cc{b} shows the selected event (b1) and its similar segments (b2 and b3). Experts are able to quickly assign tags to them one after another, or even all at once (Fig.~\ref{fig:side_panels}\cc{c} --- ``override all segments tags'').

\textbf{Decision support.} 
Sometimes the status of an anomaly is ambiguous, and choosing a proper tag is difficult. 
In this case, an expert can use shape-matching to find similar shapes (non-anomalous segments) to this event. By comparing the differences between them, experts can quickly clarify why the anomaly detection algorithm identified this segment as anomalous. 
We introduce a superposed visualization in Section~\ref{sec:similar_segment_view} that allows for such comparisons. Fig.~\ref{fig:side_panels}\cc{c} shows one example, where the event under investigation can be compared to similar segments.  

\textbf{False alarm mitigation.}
After events have been properly annotated, we are able to mitigate false alarms (false positives) by leveraging the shape-matching algorithm. 
Consider an event that experts may think is unworthy of exploration --- say, one tagged as ``Do not investigate'' or ``Normal'' (see details in Section.~\ref{sec:event_detail_view}). We can ask the system to prune existing events similar to this one using the method described below:

We use the shape of one such event as a template to search a set of candidate similar shapes, which can be tuned according to a pre-defined threshold. Those supposedly anomalous segments identified by the ML pipeline are now pruned by checking whether they overlap with these candidate similar shapes, thus reducing the number of false positives.
%A similar strategy can be applied to mitigate false negatives by leveraging ``problem'' or ``previously seen'' tagged events.

%!TEX root = ../main.tex
% !TeX spellcheck = en_US

\subsection{Visualization and Interaction}
\label{sec:mtv_visualization}
\label{sec:vis}
Our interface consists of two pages: the Landing Page and the Investigating Page.
An expert starts on the Landing Page (Fig.~\ref{fig:landing}), which provides an overview of anomaly detection results and allows experts to select certain signals and ML results for further investigation.
The Investigating Page (Fig.~\ref{fig:teaser}), MTV's most important component, contains three major panels: the Signal Overview \cc{a}, the Signal Focused View \cc{b}, and the Side Panel \cc{c}, with four collapsible sub-views included.
All three panel views work in a synchronized fashion to support the workflow described in Table~\ref{tb:workflow}.

\subsubsection{\textbf{Landing Page}}
\label{sec:landing_page}

Given a time series dataset consisting of many signals, a machine learning expert will explore a collection of ML pipelines and hyperparameter settings to identify anomalies.
The results of these pipelines are then stored in the database (\textbf{T1}). 
However, an individual or team of experts may be interested in monitoring only a subset of signals and results. 
This selection is often driven by the experts' particular expertise and the problems they're interested in tackling.
The Landing Page is designed to offer them this flexibility (\textbf{T2}).

\begin{figure}[!htbp]
	\centering
	\includegraphics[width=1.0\linewidth]{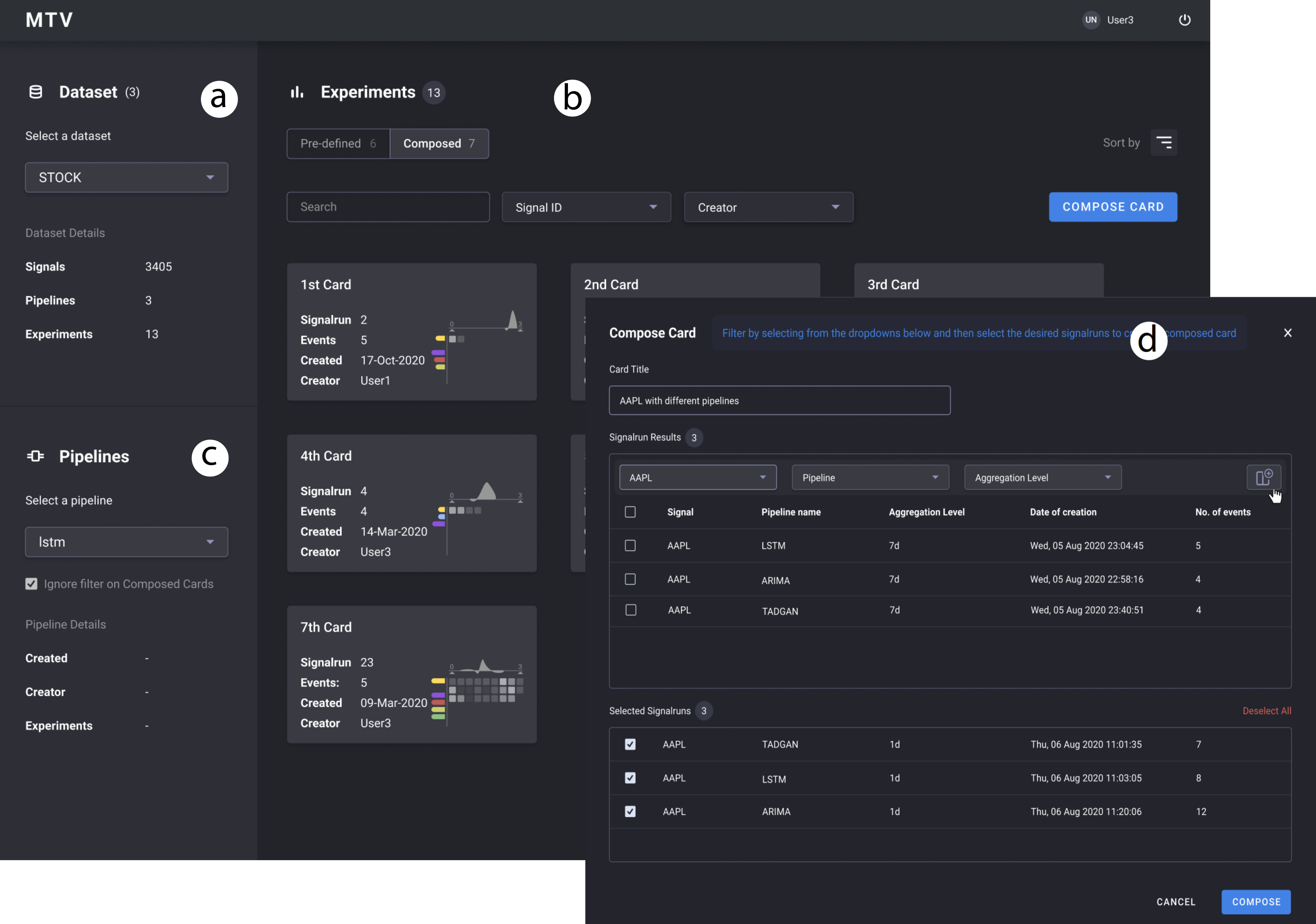}
	\caption{The Landing Page. \revise{The Dataset Panel (a) allows users to select a dataset and observe its basic information. The Main Dashboard (b) lists all the experiment cards following a user-determined order and can be filtered by whether certain pipelines have been applied in the Pipeline Panel (c). Users can click the ``COMPOSE CARD'' button to enter the Compose Card Window (d) where they can select ML outputs of interest to create a new experiment card for future collaborative analysis.}}
	\label{fig:landing}
\end{figure}

 An expert can choose different signal datasets to explore. In Fig.~\ref{fig:landing}\cc{a}, we see that the STOCK dataset contains 3405 signals (i.e., stocks), 3 pipelines (i.e., ARIAM, TADGAN, and LSTM) and 13 experiments. An experiment is defined as the outputs of one ML pipeline on a certain signal subset.

\begin{wrapfigure}{r}{0.2\textwidth}
  \vspace{-0.5cm}
  \begin{center}
    \includegraphics[width=0.19\textwidth]{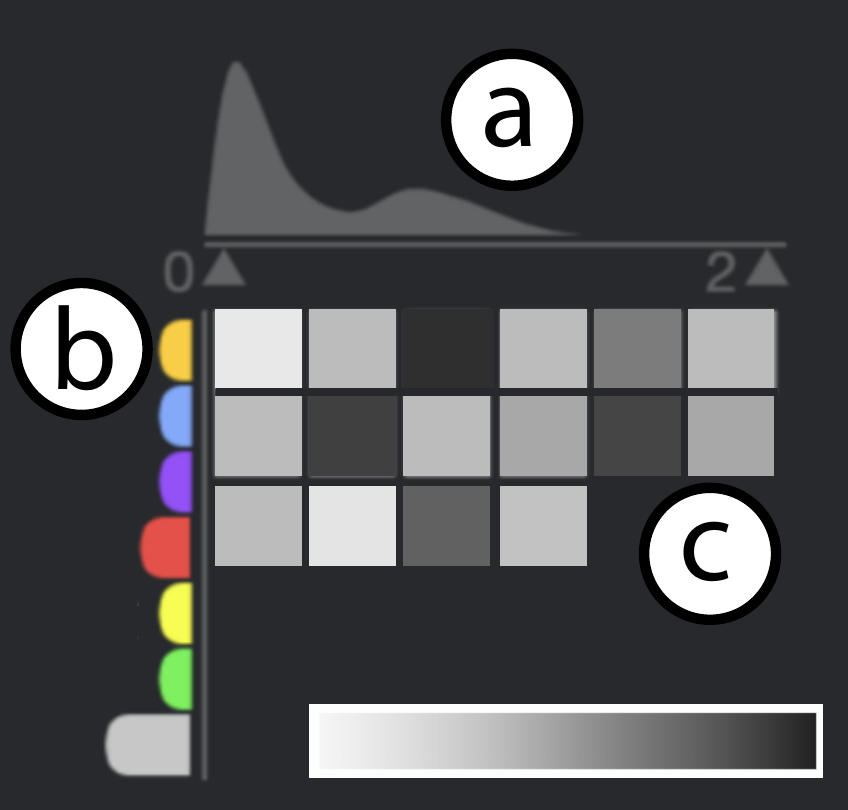}
  \end{center}
  \vspace{-0.5cm}
\end{wrapfigure}
We propose a novel card-style visualization to display the summary information for each experiment. The ``experiment card'' shows three valuable pieces of information at a glance (shown on the right).
The area chart \cc{a} at the top encodes the distribution of severity scores according to (Eq.~\ref{eq:score}). 
The colored bar chart \cc{b} on the left side shows the number of events and their corresponding tags (Sec.~\ref{sec:event_detail_view}).
By checking the information encoded in the previous two charts, it is possible to obtain quick insights about the overall severity and urgency of this analysis task and track its progress.
The matrix \cc{c} offers experts information about how anomalies are distributed over signals. Each cell of the matrix represents one signal. The cell color encodes the number of anomalies detected in that signal, and a darker shade indicates a larger number of detected anomalies.

All experiment cards are listed in the right area of the Landing Page (Fig.~\ref{fig:landing}\cc{b}) following a user-determined order. Various filter types, sorting by signal IDs, creator and/or card title, can be applied to facilitate efficient card exploration. In particular, the cards can be filtered by whether certain pipelines have been applied (Fig.~\ref{fig:landing}\cc{c}).

To provide further flexibility, the Landing Page also allows experts to select their ML pipeline and hyperparameter settings by exposing a minimal set of options to compose a new experiment card.
These options are intuitive and easily understandable. They include several important parameters such as the ability to select the level of aggregation, the strategy for imputing a missing value, or a particular ML algorithm. Fig.~\ref{fig:landing}\cc{d} shows an example experiment card, created by an expert and titled ``AAPL with different pipelines.'' This card is meant to compare the difference in results when running three different pipelines on AAPL (Apple) stock, with an aggregation level of one day. 

Finally, we stress the importance of the Landing Page, as it is an essential step (\textbf{T2}) of the workflow for general time series anomaly analysis. However, the design of this page could be domain-specific. 
Fig.~\ref{fig:landing} shows a prototype version that we implemented for experimental purposes.
However, for different application scenarios, the names of UI components in Fig.~\ref{fig:landing} could be changed, and the set of filtering options in \cc{d} varied.

% \textbf{Example Use Case}: Suppose Jane is an expert specializing in satellite data. 
% Her job is to detect potential system failures due to overheating and subsequent loss of components. 
% Jane wants to investigate anomalies in the thermal control systems (TCS).
% Thus, she creates an ``analysis card'' with the corresponding TCS signals. 
% The ``analysis card'' contains three valuable information at a glace (shown on the right).
% The area chart on the top (a) encodes the distribution of severity scores according to (Eq.~\ref{eq:score}). 
% The colored bar chart (b) attached to the left side shows the number of events and their corresponding tags (Sec.~\ref{sec:event_detail_view}).
% The block-like chart (c) where each cell represents one signal in TCS. The cell color encodes the number of anomalies detected in that signal, in which a darker color indicates a larger number of detected anomalies.  
% By assigning a name to this ``analysis card'', Jane can pull it up at any time in order to obtain quick insights about the overall severity/urgency of this analysis task and track its progress.

\subsubsection{\textbf{Signal Overview}}
\label{sec:signal_overview}
Once experts select their ``experiment card'', they go to the Investigating Page for further analysis (Fig.~\ref{fig:teaser}).
The first view in the Investigating Page is the Signal Overview. 
This view (Fig.~\ref{fig:teaser}\cc{a}) presents experts with an efficient way to scan the dynamics of every signal, as well as how anomalies are distributed.
This helps experts make informed decisions about which signals to investigate further (\textbf{T3}).

Given limited screen space, the design of multivariate time series should be highly space-efficient. 
Therefore, we choose \textit{small multiples}, a space-efficient technique that is widely used for visualizing multiple time series~\cite{javed2010graphical}.
The timeline of each signal is aligned horizontally.
In addition, we opt to use a line chart rather than another chart type (e.g., area chart, horizon graph), 
because of its interpretability. 
%this sentence below is hard to parse: kalyan
Line charts are well-suited for point-wise inspection. Additionally, line charts make trend lines very apparent, making it possible to visualize anomalies through simple trend tracking~\cite{javed2010graphical}. 
We visualize anomalous events by highlighting the curves with a warning color (Fig.~\ref{fig:teaser}-a1). 
Events across signals that appear close together in time, known as co-occurring patterns, may indicate a larger event. This further suggests that the signal order has a significant impact on experts' ability to investigate. 
Thus, we propose a novel layout algorithm to optimize the order.

\textbf{Order optimization:}
Our goal is to put ``similar'' signals as close together as possible. Similarity here refers to the length of overlapping anomalous events present in two signals.
To that end, we present a novel metric to measure the similarity between two signals $\mathbf{x}_a$ and $\mathbf{x}_b$:
\begin{equation}
Sim_{\mathbf{x}_a, \mathbf{x}_b} = \frac{|\mathbf{A}_a \cap \mathbf{A}_b|}{\min(|\mathbf{A}_a|, |\mathbf{A}_b|)}
\label{eq:similarity}
\end{equation}
where $\mathbf{A}_a$ and $\mathbf{A}_b$ are the sets of anomalous events detected in the first and second signal respectively, generally represented as $\mathbf{A} = \{(t_s, t_e)^i \: |\: i = 1, \dots, K\}$ \revise{with $K$ denoting the number of identified anomalous events in a particular signal.}
In addition, $|\mathbf{A}| = \sum_{k=1}^K (\hat{t}_e - \hat{t}_s)$ denotes the total length of the events. 
The intersection of the two event sets, denoted by $\mathbf{A}_a \cap \mathbf{A}_b$, is the event set containing all timestamps from $\mathbf{A}_a$ that also belong to $\mathbf{A}_b$. For example, when $\mathbf{A}_a = \{(1, 5), (8, 12)\}$ and $\mathbf{A}_b = \{(4, 6), (9, 11)\}$, their intersection should be $\{(4, 5), (9, 11)\}$ whose length is $1+2=3$.

This metric is inspired by the \textit{Overlap Coefficient}~\cite{vijaymeena2016survey}, which measures the overlap between two finite sets, defined by the equation $overlap(X, Y) = \frac{|X \cap Y|}{\min(|X|, |Y|)}$. It is known that the Jaccard Similarity becomes inefficient when two sets vary significantly in size, but the Overlap Coefficient overcomes this issue. If set $X$ is a subset of $Y$, or the converse, the overlap coefficient is equal to 1.
Our defined metric $Sim_{\mathbf{x}_a, \mathbf{x}_b}$ shares the same advantages.

After defining the similarity metric between signals, we can now use dimensionality reduction techniques to put the signals in sequence. 
% to map all the signals onto 1D space
In our case, we choose t-distributed stochastic neighbor embedding (T-SNE~\cite{maaten2008visualizing}) to represent all signals as single-value embedded codes. The signals are then sorted from top to bottom in the Signal Overview by their code values. 
% The algorithm is proven effective to preserve local structures which matches the fact that experts can only simultaneously observe several signals.

% \textbf{Example Use Case}: Fig.~\ref{fig:teaser}
% XX

\subsubsection{\textbf{Signal Focus View}}
\label{sec:signal_focus}
This view (Fig.~\ref{fig:teaser}\cc{b}) extends a piece of the timeline from the selected signal in the Signal Overview, displaying more information (\textbf{T4}). 
Experts can either mouse over context charts (Fig.~\ref{fig:teaser}-a2) or use the zoom and pan functions in the Signal Focus View for flexible exploration.
\revise{As shown in Fig.~\ref{fig:teaser}-b2, anomalies are highlighted such that the color of the line corresponds with the color of the tag associated with the event. To further enhance awareness, the color of the header bar double-encodes this tag information.} 
In addition, a transparent grey background is added to make anomalies more visually apparent.
% experts can create a new event by brushing a new window or modify an existing event by clicking on it (\ano{d2} in Fig.~\ref{fig:interface}\cc{d}, \textbf{R6}).
Prediction results are visualized with a thinner curve in bright \mbox{yellow}.

The smoothed errors (Eq.~\ref{eq:errors}) are represented as a centered flow on the top of the chart, where \revise{a thicker flow width indicates larger discrepancies between the original values and the predicted values (Fig.~\ref{fig:teaser}-b1).}
This visualization \revise{increases the transparency of the model} and enables experts to visually evaluate the quality of the model as well as interpret why a certain anomaly was identified by the ML pipeline.
Error flows can also be used by experts as a visual clue, \revise{guiding them to interact with (CRUD) anomalies (\textbf{T5}).}
% indicating the location of some anomalies (\textbf{T5}).

\subsubsection{\textbf{Periodical View}}
This view (Fig.~\ref{fig:teaser}\cc{c}) is designed for analyzing periodic patterns of the selected signal (\textbf{T4}).
The table on the top (Fig.~\ref{fig:teaser}-c1) summarizes the number of overall tags, as well as tags per year and per month, for a signal in focus.
The bottom graph (Fig.~\ref{fig:teaser}-c2) provides experts with a new perspective for exploring the periodical patterns of a signal.
Three levels of periodical glyphs, corresponding with the three different time granularities (i.e., year, month, and day), are proposed to support \mbox{multi-scale analysis}.

\begin{wrapfigure}{r}{0.20\textwidth}
  \begin{center}
    \vspace{-0.5cm}
    \includegraphics[width=0.19\textwidth]{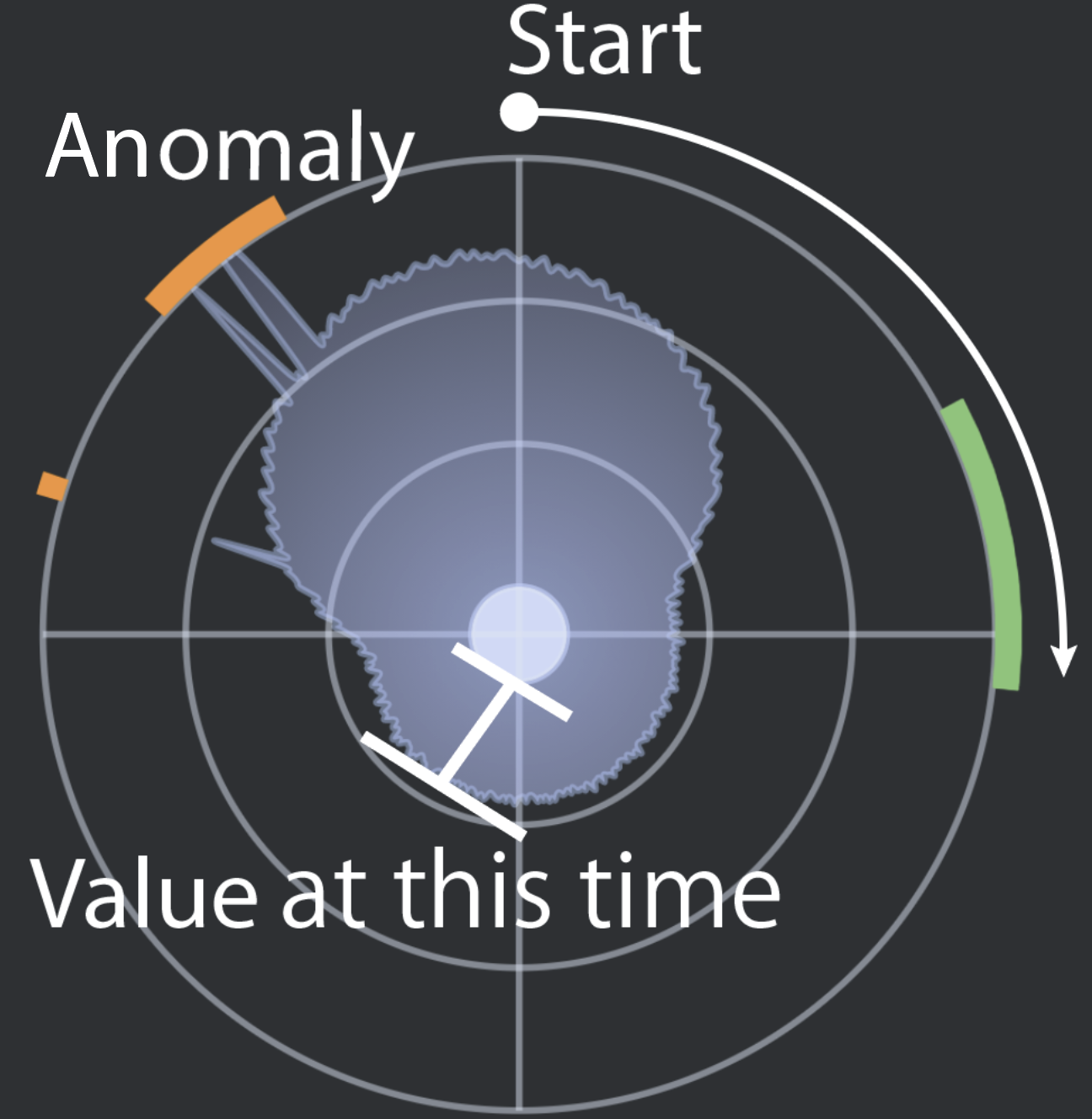}
    \vspace{-0.5cm}
  \end{center}
\end{wrapfigure}
The glyph design (shown on the right) is inspired by the circular silhouette graph~\cite{aigner2011visualization}; the glyph employs a polar coordinates system, where the angle encodes the time point in a year/month/day and the radius indicates the value at this time point.
% Circular silhouette graph is not space-efficient; thus we propose to use small multiples to support side-by-side comparison.
%kalyan note: I don't understand "small multiples"
We propose the use of \textit{small multiples}~\cite{javed2010graphical} due to its spatial efficiency and the ease of side-by-side comparison.
If for one time period (year/month/day) a signal has an irregular shape or unusual spikes, this may indicate anomalies.
In addition, we highlight the anomalies, according to their tags, by overlaying radial segments in the corresponding time periods. This enables experts to observe how anomalies are distributed across years/months/days.
For example, in Fig.~\ref{fig:teaser}-c2, the green-tagged event and red-tagged event occurred near to the start or the end of the year, while the three orange-tagged events are close to the middle of the year. 

\subsubsection{\textbf{Signal Annotations View}}
\label{sec:signal_annotation_view}
This view (Fig.~\ref{fig:side_panels}\cc{a}) provides an overview of all tags across the currently selected signal. 
From here, experts can quickly glance through the event information (starting and ending times) and what tags are associated with these events (\textbf{T4}).
We follow the design of GitHub issue labels to visualize the tag --- a rounded rectangle with the background encoding the tag type and the text showing its associated name and meaning.
Experts can click to open one event and explore the most recent annotations associated with the event. To improve efficiency, experts are allowed to directly post their comments or assign a tag here (\textbf{T5}).
These events are chronologically ordered (from top to bottom in Fig.~\ref{fig:side_panels}\cc{a}), and the sequence corresponds with the order (from left to right in Fig.~\ref{fig:teaser}\cc{b}) on the focused view, that is, green-orange-orange-orange-red.

\subsubsection{\textbf{Event Details View and Multi-Aggregation Viewer}}
\label{sec:event_detail_view}
\begin{figure}[!htbp]
	\centering
	\includegraphics[width=1\linewidth]{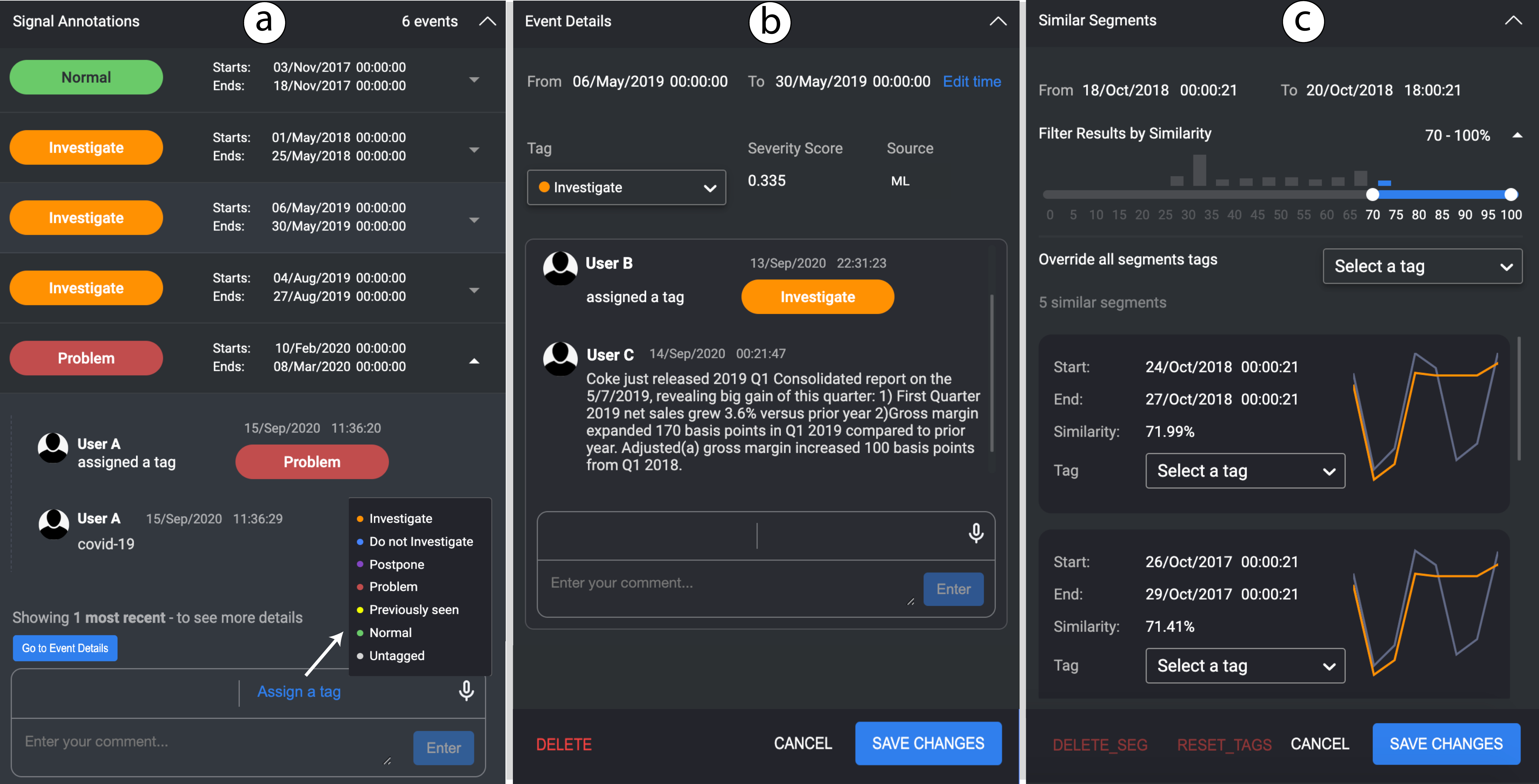}
	\caption{\revise{Three sub-views in the Side Panel: (a) the Signal Annotation View provides an overview of the annotations (ordered by event time) made for the selected signal; (b) the Event Details View shows more details about one particular event, such as severity score and source; (c) the Similar Segments View displays the search results of the shape-matching algorithm and allows users to perform quick annotations.}}
	\label{fig:side_panels}
\end{figure}

Experts can either go into the Event Details View (Fig.~\ref{fig:side_panels}\cc{b}) by directly clicking it on the Side Panel (Fig.~\ref{fig:teaser}\cc{c}) or by using the "go to Event Details" button (Fig.~\ref{fig:side_panels}\cc{a}) from the Signal Annotations View.
When the view is opened, the focal chart will be equipped with the Multi-Aggregation Viewer (Fig.~\ref{fig:6657}\cc{c}), where experts can choose different granularities to explore contextual information about the current anomaly (\textbf{T4}). 

The Event Details View, from top to bottom, displays the starting and ending times, tag information, severity score (valid only when the source is "ML"), source, and the comment box.
The source can be either "ML", "USER", or "Shape-matching."
In this view, the comment box shows all the historical annotations of the event (\textbf{T4}), in contrast to the Signal Annotations View, where only the five most recent annotations are shown.
\revise{Along with the other coordinated views, this view allows experts to perform in-situ annotation and communication (\textbf{T5}), with all the necessary contextual information displayed on one screen.}

We have designed six general types of tags, plus the status ``untagged,'' to assist collaborations between experts.
Here are these tags and their meanings:
(1) \texttt{Do not investigate} (action tag): "We are not interested in this and have decided not to investigate."  
(2) \texttt{Postpone} (action tag): "This event is interesting but of low priority; we will postpone its investigation."
(3) \texttt{Investigate} (action tag): "This event is interesting and we should investigate it now."
(4) \texttt{Problem} (info tag): 
"This is a new problem, and while we can describe it colloquially, we don't have a term for it yet." 
(5) \texttt{Previously seen} (info tag):
"This is a well-known problem that we have investigated before."
(6) \texttt{Normal} (info tag):
"This event is normal, has an obvious explanation, and is not harmful."

The tag design is the result of many design meetings with our domain experts (i.e., P1-P6 and E1-E3).
The first three action tags are meant to suggest the next step that should be taken pertaining to an event, while the last three explain what has already been decided.
\minorrevise{The tag of one event can be changed over time. In practice, if an event is tricky, experts often use action tags initially to facilitate team communication, and switch to a specific info tag later on when they reach a consensus. }

\subsubsection{\textbf{Similar Segments View}}
\label{sec:similar_segment_view}
\begin{figure}[!htbp]
	\centering
	\includegraphics[width=1\linewidth]{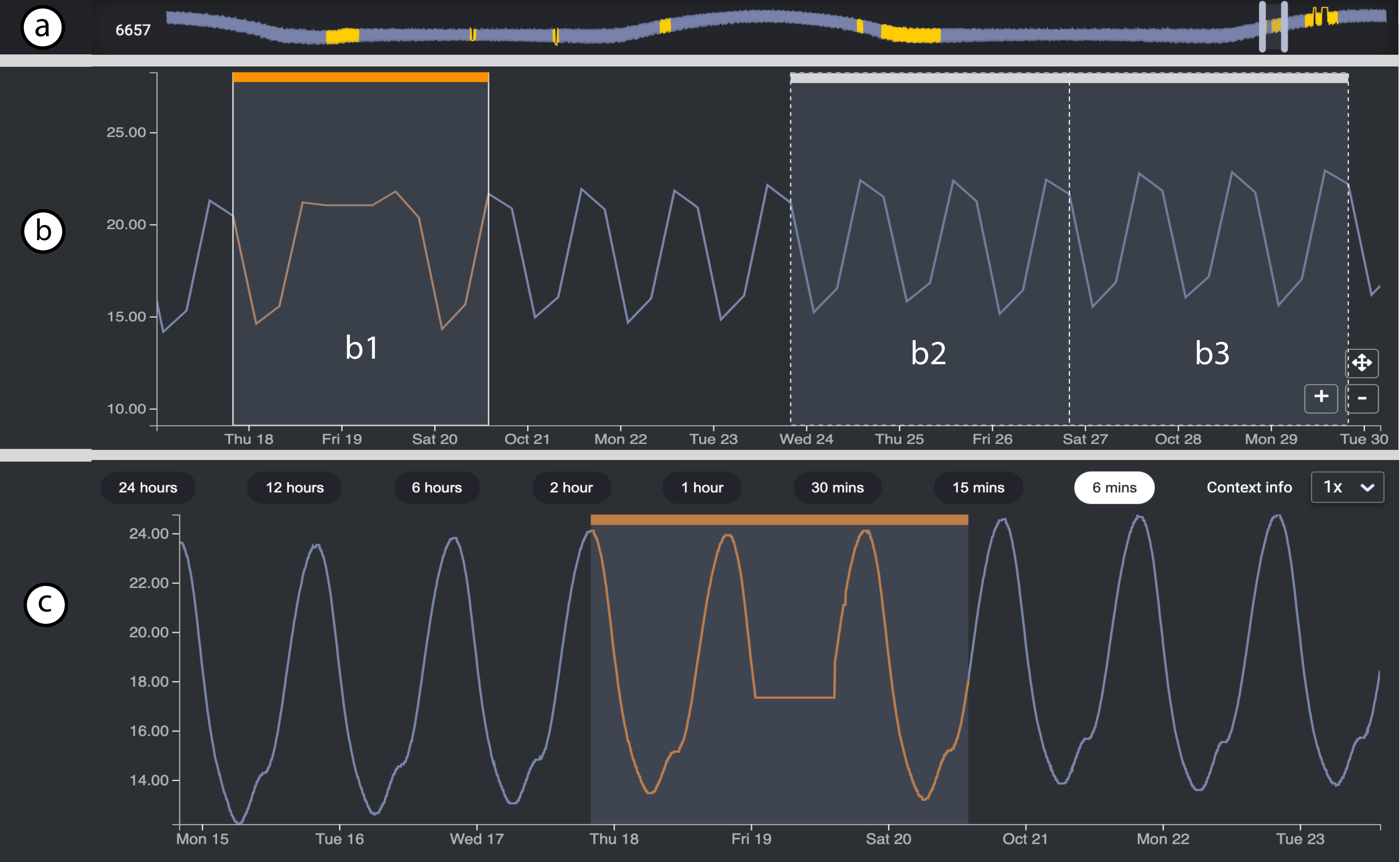}
    \caption{(a) The overall time series for signal X --- a temperature signal from a spacecraft; (b) The interval b1 is identified by ML and tagged as  ``investigate'' by one expert (P2). The expert uses the Similar Segments View (Fig.~\ref{fig:side_panels}-right) to find similar segments for comparative analysis. (c) The interval b1 is observed at a 6-minute aggregation level using the Multi-Aggregation Viewer. Seeing this, P3 suspects that this anomaly was the result of a time gap issue (missing values at a certain time period). The missing values were then filled in using the last valid value, leading to a flat line in the 6-minute level; this makes shape b1 (in the 6-hour level) stand out, showing an unusual pattern even compared with its most similar segments (b2, b3). }
    \label{fig:6657}
\end{figure}

This view (Fig.~\ref{fig:side_panels}\cc{c}) provides experts with the ability to search the most similar segments for a selected event. 
% Either DTW or Euclidean (refer to Sec.~\ref{sec:shape_matching}) can be selected as the similarity measurement.
Assume a number of similar segments (up to 100 by default) are returned by the shape-matching algorithm.
The bar chart at the top of this view is used to filter the segments based on the similarity score. 
Below the chart is a list of segments showing more detailed information, such as start and end time.
The graph on the right plots the returned segment line overlaid by the original line, color-coded by its tag (in this case orange, which is consistent with its assigned tag --- Fig.~\ref{fig:6657}-b1), allowing experts to visually compare how similar they are.
Meanwhile, the corresponding similar segments on the focal charts are highlighted using the dashed white border (Fig.~\ref{fig:6657}-b2 and -b3).
\revise{As described in Section~\ref{sec:mtv_shape_matching}, the Similar Segments View can be used for annotation sharing, decision support, and false alarm mitigation, in order to boost annotation efficiency (\textbf{T5})}.

% \begin{figure}
% \begin{minipage}{0.49\columnwidth}
% \includegraphics[width=1\linewidth]{figures/multiagg.png}
% \caption{ Empty }
% \label{fig:3}
% \end{minipage}
% \hfill  % maximize the space between the minipages
% \begin{minipage}{0.49\columnwidth}
% \includegraphics[width=1\linewidth]{figures/similar_segments.png}
% \caption{ Empty }
% \label{fig:4}
% \end{minipage}
% \end{figure}

%!TEX root = ../main.tex
% !TeX spellcheck = en_US

\section{Evaluation}\label{sec:evaluation}
\label{sec:evaluation}
To understand the usability and usefulness of \name~, we carried out two user studies based around two potential stakeholders: a domain expert, who has area expertise and analyzes anomalies as part of his or her job, and a general end-user, who does not have area expertise or anomaly analysis duties, but may still be interested in performing \minorrevise{such tasks}. 

\subsection{User Study Using Spacecraft Telemetry Data with Domain Experts}
\label{evaluation:experts}
\label{sec:evaluation_experts}

We deployed \name~ with real spacecraft telemetry data in a production environment. We carried out four case studies with our expert collaborators (P1-P6) and conducted semi-structured interviews to collect qualitative feedback about \name. 

\subsubsection{\textbf{Experimental Setup}}
We prepared the experiments using 55 real signals (measuring device electrical power, thermal temperature or attitude) tracked over a 5 year period.
We ran ML pipelines with a variety of possible settings \revise{by changing different hyperparameters that the experts wanted to control, such as the aggregation level (from 6-minute to 6-hour), imputation strategy (mean values or zero-order hold), modeling algorithm, etc.
Then we asked the experts to work together to compose four experiment cards using the Landing Page. The team would later collaborate to analyze the four experiment cards.
}

For each experiment card, the experts selected between 4 and 12 signals to analyze.
\revise{ For the first card, Case 1, four signals total were chosen from a variety of spacecraft subsystems. For Case 2, eight temperature signals were selected together with four environmental signals (e.g., sun elevation). For each of Cases 3 and 4, four attitude-control signals plus two environmental signals, and four electrical-power signals and two environmental signals were chosen. 
} 

The experts were then asked to enter \name~ within the next 24 hours and analyze these experiment cards.
\revise{During this period, they can enter \name~ anytime to create comments, check comments from other team members, and if necessary, have in-situ discussions under events of interest.}
Experts were asked to record all on-screen activities while performing these tasks, and to use the think-aloud protocol. After the 24-hour period was complete, we conducted semi-structured interviews with all the experts \revise{one by one} to collect qualitative feedback.

\subsubsection{\textbf{Results}.} 
% $\;$
\begin{table}[!h]
\begin{tabular*}{\linewidth}{@{}c @{\extracolsep{\fill}} cccc@{}}
\toprule
\textbf{Case} & \textbf{ML Event} & \textbf{User-created Event} & \textbf{Comment (avg.)} & \textbf{Tag (avg.)} \\ \midrule
1                               & 38                                          & 15                                                     & 162 (3.1)                                          & 58 (1.1)                                                  \\ 
2                               & 45                                           & 12                                                     & 87 (1.5)                                          & 60 (1.1)                                                   \\ 
3                               & 40                                           & 8                                                     & 96 (2.0)                                 & 48 (1.0)                                                   \\ 
4                               & 23                                          & 10                                                     & 66 (2.0)                          & 40 (1.2)                                                   \\ \bottomrule
\end{tabular*}
\caption{ \label{tb:evtnum}
Statistics displaying the event numbers and collected annotations for all four case studies. (avg.) indicates the average annotation number per event.
}
\end{table}

\begin{figure}[!htbp]
	\centering
	\includegraphics[width=1\linewidth]{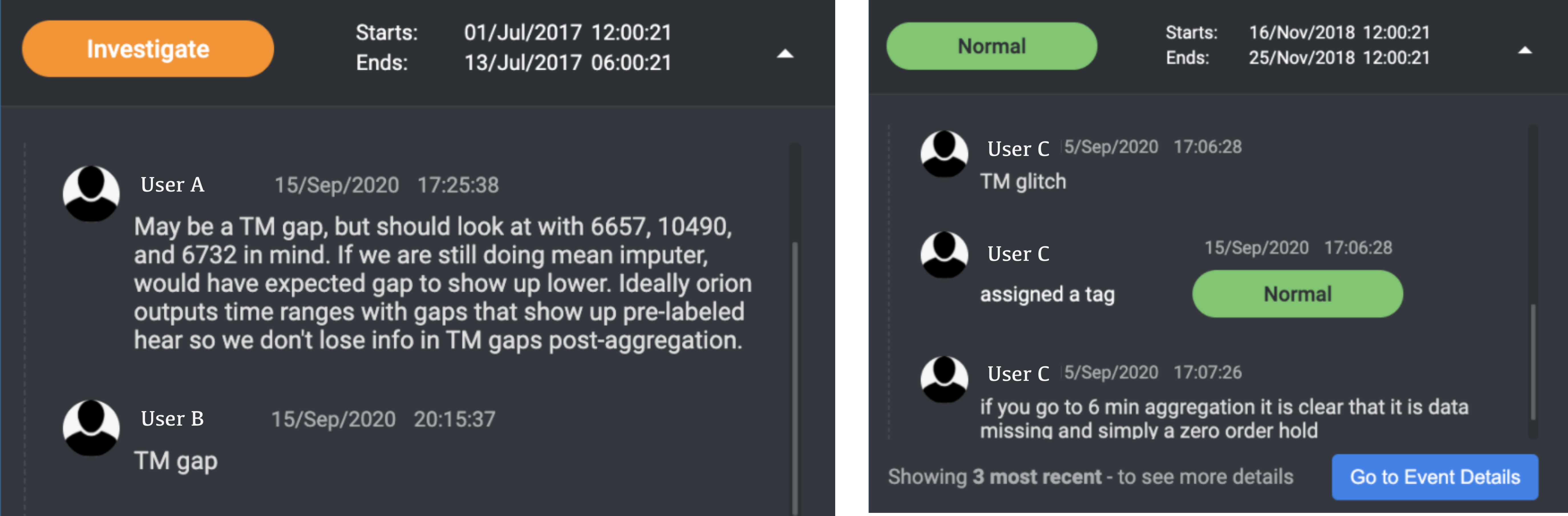}
	\caption{\revise{Example comments demonstrating how satellite experts discussed anomalous events. (Left) User A presented his suspicion that a certain anomaly may have come from a TM gap (missing values). Several hours later, User B (a more senior engineer) confirmed this suspicion and changed the tag from ``investigate'' to ``normal.'' (Right) User C, who felt confident about his finding, directly tagged the event as ``normal'' and added an explanation.}}
	\label{fig:case_ses_comments}
\end{figure}

We summarized event and annotation numbers for the four case studies in Table~\ref{tb:evtnum}.
We observed that the experts were able to used their domain knowledge to mark their own suspicious events in addition to those identified by ML. Over the four cases they created 15, 12, 8, and 10 events, respectively.

A large number of comments about one event suggests an active discussion. We observed that the average comment numbers for the four case studies were 3.1, 1.5, 2.0, and 2.0 respectively, indicating that experts can use the system to efficiently perform annotations.

Experts can perform collaborative analysis effectively as well.
Fig.~\ref{fig:6657} demonstrates the high quality of annotations through an example in which P2 investigates a temperature signal X. P2 selected a region in signal X (Fig.~\ref{fig:6657}\cc{a}) and tagged the event (b1) as ``investigate.'' 
He used the Similar Segments View to find similar segments (b2 and b3) for comparative analysis and obtained an initial suspicion: ``\textit{Maybe a time gap issue. We should look at its correlated signal Y and Z to double confirm.}'' After this comment, another expert (P3) responded, ``\textit{Signal Y and Z are fine. I think it should be a time gap issue. If go to 6-minute aggregation (Fig.~\ref{fig:6657}\cc{c}), you will find the flat line which actually is caused by missing value filling strategy}''. Finally, P4 concluded, ``\textit{Missing data issue confirmed. I will switch the current tag from investigate to normal.}''
The discussion above demonstrates that experts are able to use \name~ to conduct collaborative analysis. 
\revise{Fig.~\ref{fig:case_ses_comments} shows two additional examples that demonstrate two different types of collaborating patterns.}

\subsubsection{\textbf{Feedback Summary}.}
In general, the experts greatly appreciated the system and valued its potential to enhance their efficiency for time series anomaly analysis. They were happy to introduce the current version of \name~ to their colleagues for further use and testing. We summarize their feedback and suggestions as follows.

\textit{\textbf{Knowledge}}. P1 and P2 highlighted, ``\textit{\name~provides a comprehensive view of telemetry data. It creates a well-defined knowledge base that serves as a reference for the entire team during their investigation process.}'' Both are interested in learning how to better organize and leverage \mbox{existing annotations.}

\textit{\textbf{Scalability}}. All experts agreed that \name~ makes time series anomaly analysis possible for large-scale time series data. P4 commented, ``\textit{\name~ saves me a huge amount of effort on finding suspicious anomalies. Without the system, we are only able to monitor few numbers of signals.}'' P6 added, ``\textit{\name~ offers us a fantastic way to share and communicate what we found, before which we wasted too much time in using CSV to do the sharing.}''

\textit{\textbf{Usability}}. All the experts agreed that functionalities provided in \name~allow for efficient event exploration and effective decision making. P1 suggested, ``\textit{The error river chart is useful for me to get a sense of how serious a certain event is predicted to be. But there will be a case when all the other events are minor in comparison and cannot be seen. I suggest allowing a zoom function along y axis of this chart.}''. P6 commented, ``\textit{The Similar Segments View opens me to a brand-new way to \mbox{investigate anomalies.}}''

\textit{\textbf{Confidence}}. \name~facilitates discussions and allows the team to share insights and conduct collaborative analysis. P3 was excited about this feature, ``\textit{When I can document and share what I thought in such an organized way, I would gain more confidence in my annotations.}''

\subsection{User Study Using Stock Data with General End-Users}
\label{sec:evaluation_general}
To obtain a more comprehensive understanding of how general end-users perceive~\name, we conducted experiments using stock price data.
We selected 10 stocks from different sectors, such as energy, technology and finance, and used their daily price data since 2015. Hence, each signal (i.e. stock) contains around $2,000$ data points.

\subsubsection{\textbf{Participants.}}
We recruited 25 participants (18 male, 7 female; aged 23-40) via email invitations and on-campus advertising at 3 universities.
We did not set many constraints when selecting participants.
Each participant had between 0 and 15 years of data analytics experience ($\mu$=4.52, $\sigma$=4.59) and between 0 and 10 years of machine learning experience ($\mu$=2.42, $\sigma$=2.46).
They also had varying occupations, from students and consultants to data analysts and UI/UX designers. 
All showed a strong interest in analyzing time series data from their daily lives.
Only eight participants out of 25 had experience in the stock market. 
Because real-world users of \name~ will likely have similarly diverse backgrounds, we wanted to see whether our volunteers found the system both usable and useful. 

\subsubsection{\textbf{Experimental Setup.}}
We sent each participant a website link to perform the user study. The website included detailed instructions to guide them through the experiment.
The participants started the studies asynchronously in random order. The entire process was meant to last 1 hour.
At the start, we asked the participants to fill out a background information questionnaire. The study itself began with a 10-minute training, after which participants were asked to perform three exploration tasks and one case study.
During the training, each participant watched a tutorial video, performed three exploration tasks in \name~, and answered questions about each task to confirm their understanding of the core concepts and features. 
During the tasks, participants were asked to follow step-by-step instructions in order to explore data using particular system features.
After each task, they were asked to answer several questions about what they had found (such as the number of anomalies in stock A), as well as to provide feedback on whether a certain feature was easy to understand and use.
%The purpose of the first three tasks was to get the participant ready for the last task, the case study.

After completing the three exploration tasks, the participants were asked to collaboratively work on one open-ended task --- a case study --- creating annotations and adding their own interpretations in an asynchronous manner.
The case study involved investigating three stocks (COKE, REGN, and INTC) from different sectors --- consumer, healthcare, and technology --- together with the Nasdaq Composite (Fig.~\ref{fig:teaser}\cc{a}). The participants were able to access stock news websites or use search engines to help with the annotation. They could choose to annotate any number of events of interest, with no minimum.
Finally, they filled in post-study questionnaires regarding the effectiveness and the usability of our system, and went through a short (5 to 10 minute) informal interview in the form of one-on-one Zoom meeting.

% \noindent\textbf{Question design:}
% The survey form consists of three sections: Overall Experience, Features, and Annotations.  
% The post-study questionnaire , we used 14 5-point Likert-scale questions (1 = strongly disagree and 5 = strongly agree) to assess the overall functionality and usability of ~\name. The questions include (1) the effectiveness (in identifying anomalies), (2) powerfulness (of creating annotations), (3) ease of use, (4) ease of learning, (5) functionality (or utility), and (6) satisfaction.

% In the Features section, we asked the users to rate individual features, regarding how helpful they are. In the Annotations section, we asked the participants to provide their general thought on the quality of the annotations they made as well as the qualitative evaluation of the anomalies identified by the machines automatically. In addition, participants can write down what they like or dislike, and any additional comments in free text. 

\subsubsection{\textbf{Results}}

Our aim with this case study was to understand whether participants could create good annotations and make sense of ML results using \name.
Our participants took 12.7 minutes ($\sigma=7.2$) on average to finish their annotations during the case study.
27 events were identified by the ML algorithm, of which 25 (92.6\%) events were annotated (either tagged or commented).
11 additional events were manually created by users.
In total, the participants created 65 tags (1.7 per event) and 128 comments (3.4 per event).
We found that while a few participants did not create any annotations, this seeming lack of engagement was actually collaborative --- these users explained that they saw other people's annotations and thought they were reasonable, and felt no need to give extra explanations. 

To evaluate the quality of annotations, we sought help from 2 external volunteers with more than 3 years of stock market experience. Working together, we marked any tags or comments that did not make sense as ``invalid.'' After this, 93.8\% (61/65) tags and 91.4\% (117/128) of the total comments were considered valid. 
As an example, Fig.~\ref{fig:side_panels}\cc{a} and ~\ref{fig:side_panels}\cc{b} show real annotations made by users, which correspond to the two anomalies highlighted in Fig.~\ref{fig:teaser}-a1 (REGN) and -b2 (COKE) respectively. 
Interestingly, Fig.~\ref{fig:teaser}-b2 was first tagged by User B and later commented on by User C. The comment ``\textit{the rise is because that COKE just released 2019 Q1 Consolidated report on the May 7th 2019...}'' clearly explains the reason for this abnormal rise in the stock price of COKE.

In conclusion, we confirm that participants can use \name~ to create good annotations to make sense of, or even improve upon, ML results.

\begin{figure}[!tbp]
	\centering
% 	\vspace{-8pt}
	\includegraphics[width=1.0\linewidth]{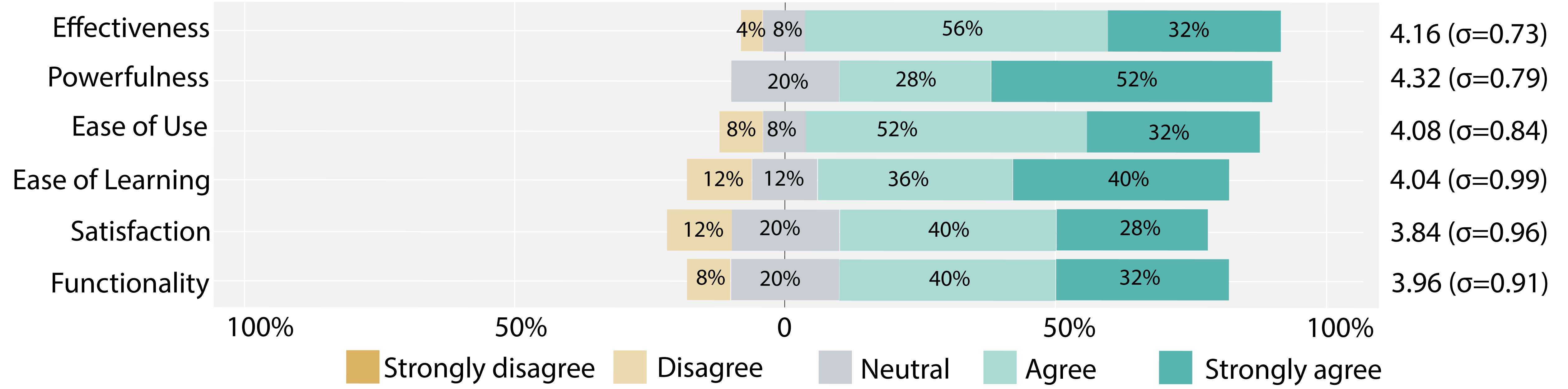}
	\caption{Ratings of the overall \name~ experience by general end-users.}
	\label{fig:result_usability}
	\vspace{-0.5cm}	
\end{figure}

\textbf{User experience --- overall assessment}.
In the post-study questionnaires, we used 5-point Likert-scale questions (1 = strongly disagree and 5 = strongly agree) to assess the overall functionality and usability of ~\name.
Fig.~\ref{fig:result_usability} shows their overall ratings. We found that in general, participants highly rated their experience with \name~.

Notably, ratings regarding usefulness are near the top of the scale, with the highest score ($\mu=4.16$, $\sigma=0.73$) for powerfulness (in annotating and sharing annotations), and the second-highest score ($\mu=4.16$, $\sigma=0.73$) for effectiveness (at detecting and investigating anomalies).
This suggests that \name~ has well achieved its goals of supporting time series anomaly detection, investigation, and collaborative annotation.
One participant (User B in Fig.~\ref{fig:side_panels}\cc{b}) commented, ``\textit{MTV is amazing tool. I love the collaboration aspect very much. I will be more confident to make my annotation when I can check other people's comments. And I felt excited when my annotation is supported by other people's comments.}''. 
Only one person (4\%, 1/25) thought our system was not effective enough to support his investigation, because he was not able to access more fine-grained price data (e.g., hourly). We think this is a flaw of the data and not directly related to our system.

The ratings for the usability aspects were slightly lower than the usefulness scores, but still good.
The participants rated ($\mu=4.08$, $\sigma=0.84$) for ease of use,  ($\mu=4.04$, $\sigma=0.99$) for ease of learning,
and ($\mu=3.84$, $\sigma=0.96$) for satisfaction (in using the system) which indicates room for further improvement.
Some participants complained about the difficulty of learning all the interaction logic for such a comprehensive system.

Users rated the functionality (a.k.a. utility) at ($\mu=3.96$, $\sigma=0.91$). A few participants (8\%, 2) thought the features provided by \name~ did not cover all their needs. For example, one wrote, ``\textit{I would like to have an option to refer to another event when adding a comment to a new one.}''. Another one said (refer to Fig.~\ref{fig:teaser}-a1), ``\textit{Looking at stock Regeneron in 2020, it has many anomalies that are nearby each other. It makes sense to merge them into one because they are caused by the Covid-19 vaccine race. So I want a function to support nearby anomaly merging}''.

\textbf{User experience --- features ratings}.
\begin{figure}[!tbp]
	\centering
	\includegraphics[width=1.0\linewidth]{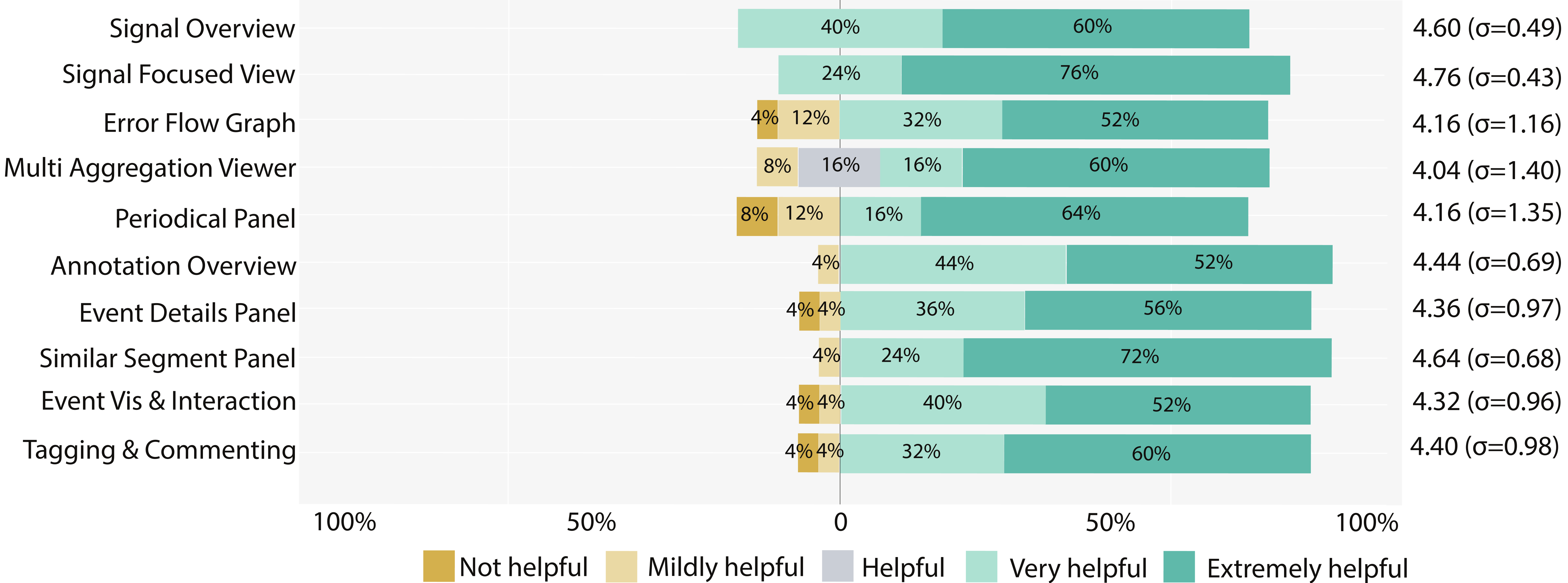}
	\caption{Ratings of individual \name~ features by general end-users.}
	\label{fig:result_feature}
% 	\vspace{-0.5cm}
\end{figure}
We also asked the participants to rate how helpful each feature was to the investigation and annotation process during the case study.
The ratings are listed in Fig.~\ref{fig:result_feature}.
Overall, the scores hovered around 4 --- ``very helpful'' --- which demonstrates users' contentment with the features.

The Signal Overview ($\mu=4.60$, $\sigma=0.49$) and the Focused View ($\mu=4.76$, $\sigma=0.43$), the two most important and frequently used features, were rated near the top of the scale.
All participants thought these two views were very or extremely helpful.
Nearly all participants (96\%, 24/25) agreed or strongly agreed on the usefulness of the Similar Segment Panel. One commented, ``\textit{The function of similar shape search is so novel to me and this panel is so well designed. It really helps me verify or spread my annotations efficiently!}''
Another interesting observation is that people's opinions on the Periodical Panel ($\mu=4.04$, $\sigma=1.40$) are slightly polarized, showing that insights regarding periodical patterns are valued differently by different people.

\subsection{Quantitative Evaluation of Algorithms}
\label{sec:evaluation_algorithms}

\begin{figure}[!htbp]
	\centering
	\includegraphics[width=1\linewidth]{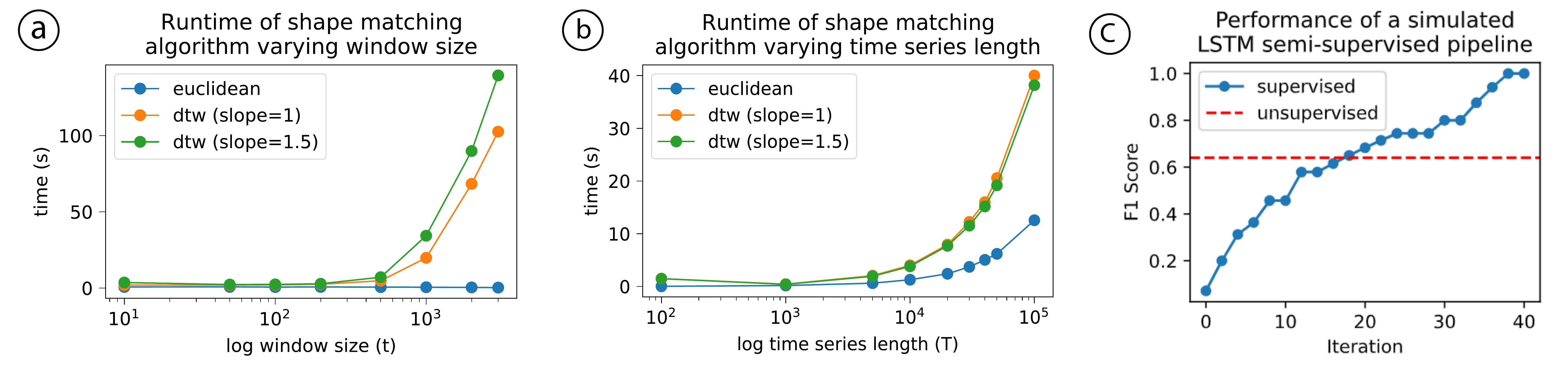}
	\caption{(a) Time performance of shape-matching with fixed time series length $5$K and varying window size (log scale); (b) Time performance of shape-matching with fixed window size $10$ and varying time series length (log scale); (c) Performance of our semi-/supervised pipeline (Fig.~\ref{fig:improve_ml}) using a simulated annotation procedure.}
	\label{fig:algorithm}
% 	\vspace{-0.5cm}
\end{figure}

\textbf{Time performance of shape-matching.}
In theory, the time complexity of Algorithm~\ref{alg:shape-matching} is $O(TD)$, where $D$ is the time needed to calculate the distance between two sub-sequences of length $t$ (i.e., the window size of the queried shape), and $T$ denotes the total length of the time series.
In the case of Euclidean distance $D$ is linear ($D = t$), while in the case of vanilla DTW $D$ is quadratic ($D = t^2$).
We consider enabling faster variations of DTW, such as the Sakoe Chiba band and the Itakura parallelogram, \minorrevise{which make the time complexity dependant on the band width and slope of the parallelogram respectively~\cite{sakoechiba, itakura}}. In real-world scenarios, we expect $t \ll T$; thus, the algorithm runs in close to linear time.

% \revise{
% \textbf{Time performance of shape-matching.}
% The novel advance of shape-matching is the ability to search for similar segments in one pass while preventing overlapping candidates, which plays a crucial role in our workflow (\textbf{T5}). The algorithm (Alg.~\ref{alg:shape-matching}) is linearly dependent on the size of the time series. While there is an additional cost involved in calculating the similarity between two sub-sequences, we observed this to be negligible in practice due to the relatively small size of the queried shape. 
% }

\revise{
Fig.~\ref{fig:algorithm}\cc{a} and ~\ref{fig:algorithm}\cc{b} report our experimental results on synthetic data when different distance metrics are applied.
\minorrevise{We generated signals of variable length $T$ of sine waves with added noise sampled from a Gaussian distribution $\mathcal{N}(0, 1)$.}
Window size denotes the number of data points of the queried shape ($t$), and time series length indicates the total length of the signal ($T$).
Note that the x-axis uses a log scale.
From Fig.~\ref{fig:algorithm}\cc{a}, we observed that the algorithm would generally take less than $2$s when the window size is smaller than $300$ and the time series length is fixed at $5$K. 
From Fig.~\ref{fig:algorithm}\cc{b}, we noted that the time cost is less than $5$s when the length of the time series is less than $10$K and the window size is fixed at $10$.
In practice, we assume the anomalous window will be small (a few to dozens of data points). Hence, our algorithm is acceptable for real-time interactions in most scenarios. }

\textbf{Human feedback evaluation.}
To validate the function of integrating human feedback back into the pipeline to improve ML performance (\textbf{T6}), we conducted a simulated experiment.
For simplicity, we assumed that experts are able to annotate $k=2$ events in a single iteration and tag an event either as ``unwanted'' (indicating that it should not be identified by ML) or ``wanted'' (indicating that it should). The simulation stopped when all events had been annotated.

A semi-supervised LSTM pipeline was trained on sequences that had already been labeled by annotators as either anomalous or normal.
We used a 70/30 data split on the NAB dataset\footnote{The Numenta Anomaly Benchmark (NAB) dataset is a well-known public time series dataset that contains 45 signals with 94 labeled anomalies. \url{https://github.com/numenta/NAB}} for training and testing.
The training data encompasses 70 events, while the test data has 32 events. Results are shown in Figure~\ref{fig:algorithm}\cc{c} where we observed that \minorrevise{the F1 score}\footnote{\minorrevise{F1 score is a measure of a model's accuracy on a dataset which ranges from 0 (the worst) to 1 (the best). }} of a semi-supervised pipeline surpasses that of an unsupervised pipeline when sufficient annotations have been obtained. In addition, observing several flat segments in the figure, we noted that some annotations may not contribute to improving detection.

%!TEX root = ../main.tex
% !TeX spellcheck = en_US

\section{Discussion and Implications}
\label{sec:discussion}

\revise{Feedback from our user studies led us to a set of design implications. We divide our discussions into three topics: visualization and interaction, machine learning, and applicability.}

\subsection{Visualization and Interaction}

\textbf{Visual scalability of the Signal Overview.}
Very occasionally, experts may compose an experiment card with many signals. Given limited screen space, the function of the Signal Overview may then be suppressed, as experts can only observe a limited number of signals at once. 
Although our proposed order optimization method can keep the ``similar'' signals close to enable better observation of co-occurring patterns, the analysis of large number of signals simultaneously can still be challenging.   
A more scalable and intuitive visual design should be considered --- one with a more information-dense design like LiveRAC~\cite{mclachlan2008liverac}, or more advanced interactions like \texttt{focus+context}~\cite{kincaid2006line}. It is worth noting that we should be careful not to increase the complexity of the original streamlined investigation workflow.

\textbf{Alternatives for visualizing multiple time series.}
\minorrevise{Providing sufficient contextual information is crucial for anomaly annotation. However, how users perceive such information depends heavily on how the time series is visualized.}
Although the line chart is our current choice for visualizing multiple time series, other visualizations, such as area charts, horizon charts, and colorfields, may also be suitable \minorrevise{in different contexts}~\cite{javed2010graphical, gogolou2018comparing}. 
% Different visual representations have their own strengths and weaknesses, some of which are contextual~\cite{javed2010graphical}.
For example, we found that the satellite experts preferred the step chart --- a variant of the line chart --- because it made it easier for them to track time gaps when a zero-order hold strategy was employed. 
In addition, a few participants in the stock experiments found it difficult to perceive nuanced differences between multiple time series. To avoid this situation, colorfields or horizon graphs might be considered as alternatives, depending on whether temporal warping or amplitude change are tolerable~\cite{gogolou2018comparing}. As future work, we will explore the best way to integrate other types of multiple time series visualization into our exploration workflow.

\textbf{Flexible tag creation.}
Our current system does not support flexible tag creation.
One obvious benefit of such a function is that experts could use a word or phrase to mark the reason for an anomaly (e.g., lunar eclipse).
However, flexible tagging also introduces additional challenges for tag management as each person may have his/her own tagging ``language''.
\minorrevise{One follow-up challenge is figuring out how to effectively utilize these tags to improve machine learning models and ensure mutually beneficial cooperation between humans and AI.}

\textbf{Anomaly comparison across signals.}
The current system does not support direct comparison of anomalies from different signals. Because experts must switch between different signals in the Signal Overview, it is also difficult for them to manually perform such comparisons. One easy solution may be the addition of a pattern screenshot function, so that when experts find useful patterns, they can save them as images for later image-level comparative analysis --- although the interactivity of this design may get lost as well. This would be an interesting direction to explore.

\subsection{Machine Learning}

\textbf{Learning from annotations.}
Currently, only info tags are fed back to the ML pipeline to enhance prediction performance.
In Section~\ref{sec:evaluation_algorithms}, we used a simulated experiment where we assumed an expert can only mark whether s/he wants a particular event to be identified in the future.
\minorrevise{
In \name's tagging system, we treat \texttt{Normal} as equivalent to ``unwanted,'' and both \texttt{Previously seen} and \texttt{Problem} as ``wanted.''     
The other three action tags (\texttt{Do not investigate}, \texttt{Postpone}, and \texttt{Investigate}) are designed to facilitate the collaborative progress of teams, and are not used in the feedback loop.
}
The semi-/supervised ML pipeline can also support multi-class classifications, but this is valuable only when the tags are more specific.
One particular challenge in regards to the learning process comes from the fact that if we depend solely on a semi-/supervised pipeline, we cannot know when it reaches the point where it performs better than an unsupervised one. 
One of our future directions involves exploring the combination of unsupervised and semi-/supervised pipelines such that they can work synchronously.

\textbf{Impact of missing values.}
\revise{
Through a series of experiments with the satellite experts, we found that many ML-identified anomalies are caused by missing values (see example in Fig.~\ref{fig:case_ses_comments}).
It is known that an unsupervised ML pipeline is only able to detect unexpected temporal patterns. A missing value is one of the most frequent reasons for a particular time segment to present an unexpected ``shape.'' Therefore, the missing value imputation strategy plays an important role in the ML pipeline. 
In future work, we want to seek the best way to encode missing value information as part of the time series visualization. Meanwhile, we plan to add hyperparameters to the ML pipeline that allow users to decide whether the pipeline should identify anomalies containing missing values. 
}

\subsection{Applicability}
\textbf{\minorrevise{Human-AI collaboration workflow} usability challenges.}
\revise{
Although the experiments described demonstrate the success of our system, the use of ML models to facilitate decision-making may introduce some potential usability challenges~\cite{zytek2021sibyl}, including lack of trust, unclear prediction targets, and difficulty reconciling human-ML disagreements. 
For instance, in our experiment with the satellite experts, $21\%$ events were still marked ``investigate'' after 24 hours because they were hard for the experts to explain.
In the long run, a growing number of`such events could introduce the aforementioned ML usability challenges.
Unlike more self-explanatory image or text data, time series data can be difficult for humans to interpret.
Although explainable AI (XAI) techniques for time series data~\cite{li2016understanding, liu2018deeptracker, schlegel2019towards, harris2020framework, cheng2021vbridge} can alleviate potential usability challenges, they may lead to cognitive biases~\cite{wang2019designing}. One promising future direction would be to formally explore which ML usability challenges exist with \name~, and investigating the best ways to integrate XAI techniques. 
}

% In addition, when experts annotate events, the system captures a variety of information about them, such as range editing, new event creation, tag changing, and concrete textual comments.
% Our current system takes advantage of certain tag information and uses a shape-matching algorithm (Sec.~\ref{sec:shape_matching}) to perform some intelligent filtering; however, there is much more space to explore such a detailed \mbox{interaction record}. 

% \textbf{Explore more usage shape-matching.}
% The novel idea behind shape-matching is how to *avoid overlapping* of candidate shapes, which our domain experts have deemed useful (R1, R2). The algorithm is linearly dependent on the size of the time series, with the additional cost of calculating the similarity between two sub-sequences. We observe that the additional cost is negligible due to the very small size of template shape. Based on our experiments, the algorithm would take 1.92s on average to compute on signals with over 31M data points.

\textbf{Use of experiment cards.}
\minorrevise{The design of experiment cards is one of the key features we propose for maintaining team awareness during collaborative analysis.}
Though we did not conduct a formal study to evaluate the usefulness of experiment cards, qualitative feedback from the satellite experts confirmed the importance of this feature.
In fact, experiment cards serve as an important bridge that connects every team member.
P1 commented ``\textit{As a manager, this feature gives me a good overlook on the progress of each currently on-going anomaly analysis task.}''.
P3 confirmed ``\textit{The design of experiment cards is not only helpful for me to track the current progress but also important for me to estimate the degree of urgency.}''.
P5 further highlighted ``\textit{The card compose function inspires me a lot of thoughts, such as comparing different algorithms' results on the same signal.}''
However, we noted that it is challenging to quickly compose a meaningful experiment card that combines as many as dozens of signals together for collaborative analysis. 
The current interactive table (Fig.~\ref{fig:landing}-d) heavily relies on domain expertise. We consider integrating relevant signal recommendation techniques as \mbox{future work.}

\textbf{Real-time analysis.}
Our current system is mainly used for historical data analysis. Depending on the domain need, the system can be updated every day or every week.
If the system is to support real-time analysis, several questions must be addressed. How does a model know if there is a shift of data distribution in streaming data? How should the model be updated if such a shift happens? What is the optimal way to update the view with incoming data and anomalies? What changes should be made in the \minorrevise{human-AI collaboration workflow} to support real-time analysis?
We aim to explore these questions as future work.

\textbf{Generalization}.
The demand for time series anomaly analysis is pervasive. Although \name~ was developed through close collaboration with spacecraft experts in particular, it (or parts of it) is generalizable to any application where large time series anomaly analysis is a crucial task --- as are the requirements for such a system, the workflow involved, and other lessons learned from the process. 
\revise{
Fig.~\ref{fig:position} clarifies the scope of applications that MTV can help with by considering decision risk, time, domain expertise, technical expertise, and the role of ML. Because the current system was designed to support asynchronous collaborative analysis, it is best suited for low or medium-risk decisions that take at least a few hours.
As future work, we are interested in exploring synchronous collaborative analysis to handle tasks that are of high decision risk and must be made in less time.
}

\section{Conclusion}\label{sec:conclusion}
We have presented ~\name, an interactive visual analysis system that allows multiple users to explore, investigate, and annotate multivariate time series collaboratively.
The system was built to facilitate a streamlined anomaly investigation workflow, which is also summarized and presented in this work.
The workflow begins with the efficient automated identification of anomalies with an end-to-end machine learning pipeline. A tailored visual interface, introduced here, allows for efficient exploration of the ML pipeline results, and features novel visualization and interaction designs that support multi-scale and multi-facet time series data exploration, as well as powerful anomaly annotation and communication.
\revise{The workflow ends by closing the loop --- training a semi-/supervised ML pipeline on collected annotations in order to enhance its performance.}
We highlight two novel algorithms --- the shape-matching algorithm and the signal layout optimization algorithm --- which we propose to better support the workflow.
\revise{We have evaluated the speed of the shape-matching algorithm, as well as the system's ability to learn from annotations.}
We have conducted user studies with two groups of people: domain experts and general end-users with an interest in analyzing time series data. Their positive feedback helps to demonstrate the effectiveness and usefulness of our system.

\begin{acks}
  We would like to thank the domain experts from SES S.A. for their invaluable insights and feedback on the system.
  We thank Sergiu Ojoc and Iulia Ionescu for their work on improving the UI/UX of MTV.
  We thank Arash Akhgari for his assistance in editing the graphics in this paper.
  We thank the participants of both user studies for their time and feedback.
  One of the co-authors Sarah Alnegheimish is supported by a scholarship from King Abdulaziz City for Science and Technology.
\end{acks}

\bibliographystyle{ACM-Reference-Format}
\bibliography{reference}


\begin{thebibliography}{78}

%%% the Web address.

\ifx \showCODEN    \undefined \def \showCODEN     #1{\unskip}     \fi
\ifx \showDOI      \undefined \def \showDOI       #1{#1}\fi
\ifx \showISBNx    \undefined \def \showISBNx     #1{\unskip}     \fi
\ifx \showISBNxiii \undefined \def \showISBNxiii  #1{\unskip}     \fi
\ifx \showISSN     \undefined \def \showISSN      #1{\unskip}     \fi
\ifx \showLCCN     \undefined \def \showLCCN      #1{\unskip}     \fi
\ifx \shownote     \undefined \def \shownote      #1{#1}          \fi
\ifx \showarticletitle \undefined \def \showarticletitle #1{#1}   \fi
\ifx \showURL      \undefined \def \showURL       {\relax}        \fi
% invisible to TeX
\providecommand\bibfield[2]{#2}
\providecommand\bibinfo[2]{#2}
\providecommand\natexlab[1]{#1}
\providecommand\showeprint[2][]{arXiv:#2}

\bibitem[\protect\citeauthoryear{Ahmad, Lavin, Purdy, and Agha}{Ahmad
  et~al\mbox{.}}{2017}]%
        {ahmad2017unsupervised}
\bibfield{author}{\bibinfo{person}{Subutai Ahmad}, \bibinfo{person}{Alexander
  Lavin}, \bibinfo{person}{Scott Purdy}, {and} \bibinfo{person}{Zuha Agha}.}
  \bibinfo{year}{2017}\natexlab{}.
\newblock \showarticletitle{Unsupervised real-time anomaly detection for
  streaming data}.
\newblock \bibinfo{journal}{\emph{Neurocomputing}}  \bibinfo{volume}{262}
  (\bibinfo{year}{2017}), \bibinfo{pages}{134--147}.
\newblock


\bibitem[\protect\citeauthoryear{Aigner, Miksch, Schumann, and Tominski}{Aigner
  et~al\mbox{.}}{2011}]%
        {aigner2011visualization}
\bibfield{author}{\bibinfo{person}{Wolfgang Aigner}, \bibinfo{person}{Silvia
  Miksch}, \bibinfo{person}{Heidrun Schumann}, {and} \bibinfo{person}{Christian
  Tominski}.} \bibinfo{year}{2011}\natexlab{}.
\newblock \bibinfo{booktitle}{\emph{Visualization of time-oriented data}}.
\newblock \bibinfo{publisher}{Springer Science \& Business Media}.
\newblock


\bibitem[\protect\citeauthoryear{Alsallakh, B{\"o}gl, Gschwandtner, Miksch,
  Esmael, Arnaout, Thonhauser, and Z{\"o}llner}{Alsallakh
  et~al\mbox{.}}{2014}]%
        {alsallakh2014visual}
\bibfield{author}{\bibinfo{person}{Bilal Alsallakh}, \bibinfo{person}{Markus
  B{\"o}gl}, \bibinfo{person}{Theresia Gschwandtner}, \bibinfo{person}{Silvia
  Miksch}, \bibinfo{person}{Bilal Esmael}, \bibinfo{person}{Arghad Arnaout},
  \bibinfo{person}{Gerhard Thonhauser}, {and} \bibinfo{person}{Philipp
  Z{\"o}llner}.} \bibinfo{year}{2014}\natexlab{}.
\newblock \showarticletitle{A visual analytics approach to segmenting and
  labeling multivariate time series data}. In \bibinfo{booktitle}{\emph{EuroVA@
  EuroVis}}.
\newblock


\bibitem[\protect\citeauthoryear{Amershi and Morris}{Amershi and
  Morris}{2008}]%
        {amershi2008cosearch}
\bibfield{author}{\bibinfo{person}{Saleema Amershi} {and}
  \bibinfo{person}{Meredith~Ringel Morris}.} \bibinfo{year}{2008}\natexlab{}.
\newblock \showarticletitle{CoSearch: a system for co-located collaborative web
  search}. In \bibinfo{booktitle}{\emph{Proceedings of the SIGCHI conference on
  human factors in computing systems}}. \bibinfo{pages}{1647--1656}.
\newblock


\bibitem[\protect\citeauthoryear{An and Cho}{An and Cho}{2015}]%
        {an2015variational}
\bibfield{author}{\bibinfo{person}{Jinwon An} {and} \bibinfo{person}{Sungzoon
  Cho}.} \bibinfo{year}{2015}\natexlab{}.
\newblock \showarticletitle{Variational autoencoder based anomaly detection
  using reconstruction probability}.
\newblock \bibinfo{journal}{\emph{Special Lecture on IE}} \bibinfo{volume}{2},
  \bibinfo{number}{1} (\bibinfo{year}{2015}), \bibinfo{pages}{1--18}.
\newblock


\bibitem[\protect\citeauthoryear{Angiulli and Pizzuti}{Angiulli and
  Pizzuti}{2002}]%
        {angiulli2002fast}
\bibfield{author}{\bibinfo{person}{Fabrizio Angiulli} {and}
  \bibinfo{person}{Clara Pizzuti}.} \bibinfo{year}{2002}\natexlab{}.
\newblock \showarticletitle{Fast outlier detection in high dimensional spaces}.
  In \bibinfo{booktitle}{\emph{European conference on principles of data mining
  and knowledge discovery}}. Springer, \bibinfo{pages}{15--27}.
\newblock


\bibitem[\protect\citeauthoryear{Bardram and Houben}{Bardram and
  Houben}{2018}]%
        {bardram2018collaborative}
\bibfield{author}{\bibinfo{person}{Jakob~E Bardram} {and}
  \bibinfo{person}{Steven Houben}.} \bibinfo{year}{2018}\natexlab{}.
\newblock \showarticletitle{Collaborative affordances of medical records}.
\newblock \bibinfo{journal}{\emph{Computer Supported Cooperative Work (CSCW)}}
  \bibinfo{volume}{27}, \bibinfo{number}{1} (\bibinfo{year}{2018}),
  \bibinfo{pages}{1--36}.
\newblock


\bibitem[\protect\citeauthoryear{Bernard, Hutter, Zeppelzauer, Fellner, and
  Sedlmair}{Bernard et~al\mbox{.}}{2017a}]%
        {bernard2017comparing}
\bibfield{author}{\bibinfo{person}{J{\"u}rgen Bernard}, \bibinfo{person}{Marco
  Hutter}, \bibinfo{person}{Matthias Zeppelzauer}, \bibinfo{person}{Dieter
  Fellner}, {and} \bibinfo{person}{Michael Sedlmair}.}
  \bibinfo{year}{2017}\natexlab{a}.
\newblock \showarticletitle{Comparing visual-interactive labeling with active
  learning: An experimental study}.
\newblock \bibinfo{journal}{\emph{IEEE TVCG}} \bibinfo{volume}{24},
  \bibinfo{number}{1} (\bibinfo{year}{2017}), \bibinfo{pages}{298--308}.
\newblock


\bibitem[\protect\citeauthoryear{Bernard, Ritter, Sessler, Zeppelzauer,
  Kohlhammer, and Fellner}{Bernard et~al\mbox{.}}{2017b}]%
        {bernard2017visual}
\bibfield{author}{\bibinfo{person}{J{\"u}rgen Bernard},
  \bibinfo{person}{Christian Ritter}, \bibinfo{person}{David Sessler},
  \bibinfo{person}{Matthias Zeppelzauer}, \bibinfo{person}{J{\"o}rn
  Kohlhammer}, {and} \bibinfo{person}{Dieter Fellner}.}
  \bibinfo{year}{2017}\natexlab{b}.
\newblock \showarticletitle{Visual-interactive similarity search for complex
  objects by example of soccer player analysis}.
\newblock \bibinfo{journal}{\emph{arXiv preprint arXiv:1703.03385}}
  (\bibinfo{year}{2017}).
\newblock


\bibitem[\protect\citeauthoryear{Bernard, Zeppelzauer, Sedlmair, and
  Aigner}{Bernard et~al\mbox{.}}{2018}]%
        {bernard2018vial}
\bibfield{author}{\bibinfo{person}{J{\"u}rgen Bernard},
  \bibinfo{person}{Matthias Zeppelzauer}, \bibinfo{person}{Michael Sedlmair},
  {and} \bibinfo{person}{Wolfgang Aigner}.} \bibinfo{year}{2018}\natexlab{}.
\newblock \showarticletitle{VIAL: a unified process for visual interactive
  labeling}.
\newblock \bibinfo{journal}{\emph{The Visual Computer}} \bibinfo{volume}{34},
  \bibinfo{number}{9} (\bibinfo{year}{2018}), \bibinfo{pages}{1189--1207}.
\newblock


\bibitem[\protect\citeauthoryear{Berndt and Clifford}{Berndt and
  Clifford}{1994}]%
        {berndt1994using}
\bibfield{author}{\bibinfo{person}{Donald~J Berndt} {and}
  \bibinfo{person}{James Clifford}.} \bibinfo{year}{1994}\natexlab{}.
\newblock \showarticletitle{Using dynamic time warping to find patterns in time
  series}. In \bibinfo{booktitle}{\emph{KDD workshop}},
  Vol.~\bibinfo{volume}{10}. Seattle, WA, USA:, \bibinfo{pages}{359--370}.
\newblock


\bibitem[\protect\citeauthoryear{Brennan, Mueller, Zelinsky, Ramakrishnan,
  Warren, and Kaufman}{Brennan et~al\mbox{.}}{2006}]%
        {brennan2006toward}
\bibfield{author}{\bibinfo{person}{Susan~E Brennan}, \bibinfo{person}{Klaus
  Mueller}, \bibinfo{person}{Greg Zelinsky}, \bibinfo{person}{IV Ramakrishnan},
  \bibinfo{person}{David~S Warren}, {and} \bibinfo{person}{Arie Kaufman}.}
  \bibinfo{year}{2006}\natexlab{}.
\newblock \showarticletitle{Toward a multi-analyst, collaborative framework for
  visual analytics}. In \bibinfo{booktitle}{\emph{2006 IEEE VAST}}.
  \bibinfo{pages}{129--136}.
\newblock


\bibitem[\protect\citeauthoryear{Breunig, Kriegel, Ng, and Sander}{Breunig
  et~al\mbox{.}}{2000}]%
        {breunig2000lof}
\bibfield{author}{\bibinfo{person}{Markus~M Breunig},
  \bibinfo{person}{Hans-Peter Kriegel}, \bibinfo{person}{Raymond~T Ng}, {and}
  \bibinfo{person}{J{\"o}rg Sander}.} \bibinfo{year}{2000}\natexlab{}.
\newblock \showarticletitle{LOF: identifying density-based local outliers}. In
  \bibinfo{booktitle}{\emph{2000 ACM SIGMOD}}. \bibinfo{pages}{93--104}.
\newblock


\bibitem[\protect\citeauthoryear{Cao, Lin, Zhu, Lin, Teng, and Wen}{Cao
  et~al\mbox{.}}{2017}]%
        {cao2017voila}
\bibfield{author}{\bibinfo{person}{Nan Cao}, \bibinfo{person}{Chaoguang Lin},
  \bibinfo{person}{Qiuhan Zhu}, \bibinfo{person}{Yu-Ru Lin},
  \bibinfo{person}{Xian Teng}, {and} \bibinfo{person}{Xidao Wen}.}
  \bibinfo{year}{2017}\natexlab{}.
\newblock \showarticletitle{Voila: Visual anomaly detection and monitoring with
  streaming spatiotemporal data}.
\newblock \bibinfo{journal}{\emph{IEEE TVCG}} \bibinfo{volume}{24},
  \bibinfo{number}{1} (\bibinfo{year}{2017}), \bibinfo{pages}{23--33}.
\newblock


\bibitem[\protect\citeauthoryear{Cao, Shi, Lin, Lu, Lin, and Lin}{Cao
  et~al\mbox{.}}{2015}]%
        {cao2015targetvue}
\bibfield{author}{\bibinfo{person}{Nan Cao}, \bibinfo{person}{Conglei Shi},
  \bibinfo{person}{Sabrina Lin}, \bibinfo{person}{Jie Lu},
  \bibinfo{person}{Yu-Ru Lin}, {and} \bibinfo{person}{Ching-Yung Lin}.}
  \bibinfo{year}{2015}\natexlab{}.
\newblock \showarticletitle{Target{Vue}: Visual analysis of anomalous user
  behaviors in online communication systems}.
\newblock \bibinfo{journal}{\emph{IEEE TVCG}} \bibinfo{volume}{22},
  \bibinfo{number}{1} (\bibinfo{year}{2015}), \bibinfo{pages}{280--289}.
\newblock


\bibitem[\protect\citeauthoryear{Chandola, Banerjee, and Kumar}{Chandola
  et~al\mbox{.}}{2009}]%
        {chandola2009anomaly}
\bibfield{author}{\bibinfo{person}{Varun Chandola}, \bibinfo{person}{Arindam
  Banerjee}, {and} \bibinfo{person}{Vipin Kumar}.}
  \bibinfo{year}{2009}\natexlab{}.
\newblock \showarticletitle{Anomaly detection: A survey}.
\newblock \bibinfo{journal}{\emph{ACM computing surveys (CSUR)}}
  \bibinfo{volume}{41}, \bibinfo{number}{3} (\bibinfo{year}{2009}),
  \bibinfo{pages}{15}.
\newblock


\bibitem[\protect\citeauthoryear{Cheng, Liu, Du, Lin, Zytek, Li, Qu, and
  Veeramachaneni}{Cheng et~al\mbox{.}}{2021}]%
        {cheng2021vbridge}
\bibfield{author}{\bibinfo{person}{Furui Cheng}, \bibinfo{person}{Dongyu Liu},
  \bibinfo{person}{Fan Du}, \bibinfo{person}{Yanna Lin},
  \bibinfo{person}{Alexandra Zytek}, \bibinfo{person}{Haomin Li},
  \bibinfo{person}{Huamin Qu}, {and} \bibinfo{person}{Kalyan Veeramachaneni}.}
  \bibinfo{year}{2021}\natexlab{}.
\newblock \showarticletitle{VBridge: Connecting the Dots Between Features and
  Data to Explain Healthcare Models}.
\newblock \bibinfo{journal}{\emph{IEEE Transactions on Visualization and
  Computer Graphics}} (\bibinfo{year}{2021}).
\newblock


\bibitem[\protect\citeauthoryear{Dema, Brereton, Cappadonna, Roe, Truskinger,
  and Zhang}{Dema et~al\mbox{.}}{2017}]%
        {dema_collaborative_2017}
\bibfield{author}{\bibinfo{person}{Tshering Dema}, \bibinfo{person}{Margot
  Brereton}, \bibinfo{person}{Jessica~L. Cappadonna}, \bibinfo{person}{Paul
  Roe}, \bibinfo{person}{Anthony Truskinger}, {and} \bibinfo{person}{Jinglan
  Zhang}.} \bibinfo{year}{2017}\natexlab{}.
\newblock \showarticletitle{Collaborative {Exploration} and {Sensemaking} of
  {Big} {Environmental} {Sound} {Data}}.
\newblock \bibinfo{journal}{\emph{Computer Supported Cooperative Work (CSCW)}}
  \bibinfo{volume}{26}, \bibinfo{number}{4-6} (\bibinfo{date}{Dec.}
  \bibinfo{year}{2017}), \bibinfo{pages}{693--731}.
\newblock
\showISSN{0925-9724, 1573-7551}


\bibitem[\protect\citeauthoryear{Fuchs, Fischer, Mansmann, Bertini, and
  Isenberg}{Fuchs et~al\mbox{.}}{2013}]%
        {fuchs2013evaluation}
\bibfield{author}{\bibinfo{person}{Johannes Fuchs}, \bibinfo{person}{Fabian
  Fischer}, \bibinfo{person}{Florian Mansmann}, \bibinfo{person}{Enrico
  Bertini}, {and} \bibinfo{person}{Petra Isenberg}.}
  \bibinfo{year}{2013}\natexlab{}.
\newblock \showarticletitle{Evaluation of alternative glyph designs for time
  series data in a small multiple setting}. In
  \bibinfo{booktitle}{\emph{Proceedings of the SIGCHI conference on human
  factors in computing systems}}. \bibinfo{pages}{3237--3246}.
\newblock


\bibitem[\protect\citeauthoryear{Gava, Spinola, Tonini, and Medina}{Gava
  et~al\mbox{.}}{2012}]%
        {gava20123c}
\bibfield{author}{\bibinfo{person}{Vagner~Luiz Gava}, \bibinfo{person}{Mauro
  de~Mesquita Spinola}, \bibinfo{person}{Antonio~Carlos Tonini}, {and}
  \bibinfo{person}{Jos{\'e}~Cardenas Medina}.} \bibinfo{year}{2012}\natexlab{}.
\newblock \showarticletitle{The 3c cooperation model applied to the classical
  requirement analysis}.
\newblock \bibinfo{journal}{\emph{JISTEM-Journal of Information Systems and
  Technology Management}} \bibinfo{volume}{9}, \bibinfo{number}{2}
  (\bibinfo{year}{2012}), \bibinfo{pages}{235--264}.
\newblock


\bibitem[\protect\citeauthoryear{Geiger, Liu, Alnegheimish, Cuesta-Infante, and
  Veeramachaneni}{Geiger et~al\mbox{.}}{2020}]%
        {geiger2020tadgan}
\bibfield{author}{\bibinfo{person}{Alexander Geiger}, \bibinfo{person}{Dongyu
  Liu}, \bibinfo{person}{Sarah Alnegheimish}, \bibinfo{person}{Alfredo
  Cuesta-Infante}, {and} \bibinfo{person}{Kalyan Veeramachaneni}.}
  \bibinfo{year}{2020}\natexlab{}.
\newblock \showarticletitle{TadGAN: Time Series Anomaly Detection Using
  Generative Adversarial Networks}.
\newblock \bibinfo{journal}{\emph{arXiv preprint arXiv:2009.07769}}
  (\bibinfo{year}{2020}).
\newblock


\bibitem[\protect\citeauthoryear{Gogolou, Tsandilas, Palpanas, and
  Bezerianos}{Gogolou et~al\mbox{.}}{2018}]%
        {gogolou2018comparing}
\bibfield{author}{\bibinfo{person}{Anna Gogolou}, \bibinfo{person}{Theophanis
  Tsandilas}, \bibinfo{person}{Themis Palpanas}, {and}
  \bibinfo{person}{Anastasia Bezerianos}.} \bibinfo{year}{2018}\natexlab{}.
\newblock \showarticletitle{Comparing similarity perception in time series
  visualizations}.
\newblock \bibinfo{journal}{\emph{IEEE TVCG}} \bibinfo{volume}{25},
  \bibinfo{number}{1} (\bibinfo{year}{2018}), \bibinfo{pages}{523--533}.
\newblock


\bibitem[\protect\citeauthoryear{Goldstein and Uchida}{Goldstein and
  Uchida}{2016}]%
        {goldstein2016comparative}
\bibfield{author}{\bibinfo{person}{Markus Goldstein} {and}
  \bibinfo{person}{Seiichi Uchida}.} \bibinfo{year}{2016}\natexlab{}.
\newblock \showarticletitle{A comparative evaluation of unsupervised anomaly
  detection algorithms for multivariate data}.
\newblock \bibinfo{journal}{\emph{PloS one}} \bibinfo{volume}{11},
  \bibinfo{number}{4} (\bibinfo{year}{2016}), \bibinfo{pages}{e0152173}.
\newblock


\bibitem[\protect\citeauthoryear{Goyal and Fussell}{Goyal and Fussell}{2016}]%
        {goyal2016effects}
\bibfield{author}{\bibinfo{person}{Nitesh Goyal} {and} \bibinfo{person}{Susan~R
  Fussell}.} \bibinfo{year}{2016}\natexlab{}.
\newblock \showarticletitle{Effects of sensemaking translucence on distributed
  collaborative analysis}. In \bibinfo{booktitle}{\emph{Proceedings of the 19th
  ACM Conference on Computer-Supported Cooperative Work \& Social Computing}}.
  \bibinfo{pages}{288--302}.
\newblock


\bibitem[\protect\citeauthoryear{Habeeb, Nasaruddin, Gani, Hashem, Ahmed, and
  Imran}{Habeeb et~al\mbox{.}}{2019}]%
        {habeeb2019real}
\bibfield{author}{\bibinfo{person}{Riyaz Ahamed~Ariyaluran Habeeb},
  \bibinfo{person}{Fariza Nasaruddin}, \bibinfo{person}{Abdullah Gani},
  \bibinfo{person}{Ibrahim Abaker~Targio Hashem}, \bibinfo{person}{Ejaz Ahmed},
  {and} \bibinfo{person}{Muhammad Imran}.} \bibinfo{year}{2019}\natexlab{}.
\newblock \showarticletitle{Real-time big data processing for anomaly
  detection: A survey}.
\newblock \bibinfo{journal}{\emph{International Journal of Information
  Management}}  \bibinfo{volume}{45} (\bibinfo{year}{2019}),
  \bibinfo{pages}{289--307}.
\newblock


\bibitem[\protect\citeauthoryear{Hajizadeh, Tory, and Leung}{Hajizadeh
  et~al\mbox{.}}{2013}]%
        {hajizadeh2013supporting}
\bibfield{author}{\bibinfo{person}{Amir~Hossein Hajizadeh},
  \bibinfo{person}{Melanie Tory}, {and} \bibinfo{person}{Rock Leung}.}
  \bibinfo{year}{2013}\natexlab{}.
\newblock \showarticletitle{Supporting awareness through collaborative brushing
  and linking of tabular data}.
\newblock \bibinfo{journal}{\emph{IEEE TVCG}} \bibinfo{volume}{19},
  \bibinfo{number}{12} (\bibinfo{year}{2013}), \bibinfo{pages}{2189--2197}.
\newblock


\bibitem[\protect\citeauthoryear{Harris, Chen, and Zaki}{Harris
  et~al\mbox{.}}{2020}]%
        {harris2020framework}
\bibfield{author}{\bibinfo{person}{Jonathan~J Harris},
  \bibinfo{person}{Ching-Hua Chen}, {and} \bibinfo{person}{Mohammed~J Zaki}.}
  \bibinfo{year}{2020}\natexlab{}.
\newblock \showarticletitle{A Framework for Generating Explanations from
  Temporal Personal Health Data}.
\newblock \bibinfo{journal}{\emph{arXiv preprint arXiv:2003.09530}}
  (\bibinfo{year}{2020}).
\newblock


\bibitem[\protect\citeauthoryear{Heer and Agrawala}{Heer and Agrawala}{2008}]%
        {heer2008design}
\bibfield{author}{\bibinfo{person}{Jeffrey Heer} {and} \bibinfo{person}{Maneesh
  Agrawala}.} \bibinfo{year}{2008}\natexlab{}.
\newblock \showarticletitle{Design considerations for collaborative visual
  analytics}.
\newblock \bibinfo{journal}{\emph{Information visualization}}
  \bibinfo{volume}{7}, \bibinfo{number}{1} (\bibinfo{year}{2008}),
  \bibinfo{pages}{49--62}.
\newblock


\bibitem[\protect\citeauthoryear{Heer, Van~Ham, Carpendale, Weaver, and
  Isenberg}{Heer et~al\mbox{.}}{2008}]%
        {heer2008creation}
\bibfield{author}{\bibinfo{person}{Jeffrey Heer}, \bibinfo{person}{Frank
  Van~Ham}, \bibinfo{person}{Sheelagh Carpendale}, \bibinfo{person}{Chris
  Weaver}, {and} \bibinfo{person}{Petra Isenberg}.}
  \bibinfo{year}{2008}\natexlab{}.
\newblock \showarticletitle{Creation and collaboration: Engaging new audiences
  for information visualization}.
\newblock In \bibinfo{booktitle}{\emph{Information visualization}}.
  \bibinfo{publisher}{Springer}, \bibinfo{pages}{92--133}.
\newblock


\bibitem[\protect\citeauthoryear{Heer, Vi{\'e}gas, and Wattenberg}{Heer
  et~al\mbox{.}}{2007}]%
        {heer2007voyagers}
\bibfield{author}{\bibinfo{person}{Jeffrey Heer}, \bibinfo{person}{Fernanda~B
  Vi{\'e}gas}, {and} \bibinfo{person}{Martin Wattenberg}.}
  \bibinfo{year}{2007}\natexlab{}.
\newblock \showarticletitle{Voyagers and voyeurs: supporting asynchronous
  collaborative information visualization}. In
  \bibinfo{booktitle}{\emph{Proceedings of the SIGCHI conference on Human
  factors in computing systems}}. \bibinfo{pages}{1029--1038}.
\newblock


\bibitem[\protect\citeauthoryear{Heimerl, Koch, Bosch, and Ertl}{Heimerl
  et~al\mbox{.}}{2012}]%
        {heimerl2012visual}
\bibfield{author}{\bibinfo{person}{Florian Heimerl}, \bibinfo{person}{Steffen
  Koch}, \bibinfo{person}{Harald Bosch}, {and} \bibinfo{person}{Thomas Ertl}.}
  \bibinfo{year}{2012}\natexlab{}.
\newblock \showarticletitle{Visual classifier training for text document
  retrieval}.
\newblock \bibinfo{journal}{\emph{IEEE TVCG}} \bibinfo{volume}{18},
  \bibinfo{number}{12} (\bibinfo{year}{2012}), \bibinfo{pages}{2839--2848}.
\newblock


\bibitem[\protect\citeauthoryear{Hodge and Austin}{Hodge and Austin}{2004}]%
        {hodge2004survey}
\bibfield{author}{\bibinfo{person}{Victoria Hodge} {and} \bibinfo{person}{Jim
  Austin}.} \bibinfo{year}{2004}\natexlab{}.
\newblock \showarticletitle{A survey of outlier detection methodologies}.
\newblock \bibinfo{journal}{\emph{Artificial intelligence review}}
  \bibinfo{volume}{22}, \bibinfo{number}{2} (\bibinfo{year}{2004}),
  \bibinfo{pages}{85--126}.
\newblock


\bibitem[\protect\citeauthoryear{H{\"o}ferlin, Netzel, H{\"o}ferlin, Weiskopf,
  and Heidemann}{H{\"o}ferlin et~al\mbox{.}}{2012}]%
        {hoferlin2012inter}
\bibfield{author}{\bibinfo{person}{Benjamin H{\"o}ferlin},
  \bibinfo{person}{Rudolf Netzel}, \bibinfo{person}{Markus H{\"o}ferlin},
  \bibinfo{person}{Daniel Weiskopf}, {and} \bibinfo{person}{Gunther
  Heidemann}.} \bibinfo{year}{2012}\natexlab{}.
\newblock \showarticletitle{Inter-active learning of ad-hoc classifiers for
  video visual analytics}. In \bibinfo{booktitle}{\emph{2012 IEEE VAST}}.
  \bibinfo{pages}{23--32}.
\newblock


\bibitem[\protect\citeauthoryear{Hong, Suh, Henry~Riche, Lee, Kim, and
  Zachry}{Hong et~al\mbox{.}}{2018}]%
        {hong2018collaborative}
\bibfield{author}{\bibinfo{person}{Sungsoo Hong}, \bibinfo{person}{Minhyang
  Suh}, \bibinfo{person}{Nathalie Henry~Riche}, \bibinfo{person}{Jooyoung Lee},
  \bibinfo{person}{Juho Kim}, {and} \bibinfo{person}{Mark Zachry}.}
  \bibinfo{year}{2018}\natexlab{}.
\newblock \showarticletitle{Collaborative dynamic queries: Supporting
  distributed small group decision-making}. In
  \bibinfo{booktitle}{\emph{Proceedings of the 2018 CHI Conference on Human
  Factors in Computing Systems}}. \bibinfo{pages}{1--12}.
\newblock


\bibitem[\protect\citeauthoryear{Hong, Suh, Kim, Smoke, Sien, Ng, Zachry, and
  Kim}{Hong et~al\mbox{.}}{2019}]%
        {hong2019design}
\bibfield{author}{\bibinfo{person}{Sungsoo Hong}, \bibinfo{person}{Minhyang
  Suh}, \bibinfo{person}{Tae~Soo Kim}, \bibinfo{person}{Irina Smoke},
  \bibinfo{person}{Sangwha Sien}, \bibinfo{person}{Janet Ng},
  \bibinfo{person}{Mark Zachry}, {and} \bibinfo{person}{Juho Kim}.}
  \bibinfo{year}{2019}\natexlab{}.
\newblock \showarticletitle{Design for Collaborative Information-Seeking:
  Understanding User Challenges and Deploying Collaborative Dynamic Queries}.
\newblock \bibinfo{journal}{\emph{Proceedings of the ACM on Human-Computer
  Interaction}} \bibinfo{volume}{3}, \bibinfo{number}{CSCW}
  (\bibinfo{year}{2019}), \bibinfo{pages}{1--24}.
\newblock


\bibitem[\protect\citeauthoryear{Hundman, Constantinou, Laporte, Colwell, and
  Soderstrom}{Hundman et~al\mbox{.}}{2018}]%
        {hundman2018detecting}
\bibfield{author}{\bibinfo{person}{Kyle Hundman}, \bibinfo{person}{Valentino
  Constantinou}, \bibinfo{person}{Christopher Laporte}, \bibinfo{person}{Ian
  Colwell}, {and} \bibinfo{person}{Tom Soderstrom}.}
  \bibinfo{year}{2018}\natexlab{}.
\newblock \showarticletitle{Detecting spacecraft anomalies using lstms and
  nonparametric dynamic thresholding}. In \bibinfo{booktitle}{\emph{Proceedings
  of the 24th ACM SIGKDD International Conference on Knowledge Discovery \&
  Data Mining}}. ACM, \bibinfo{pages}{387--395}.
\newblock


\bibitem[\protect\citeauthoryear{Isenberg and Carpendale}{Isenberg and
  Carpendale}{2007}]%
        {isenberg2007interactive}
\bibfield{author}{\bibinfo{person}{Petra Isenberg} {and}
  \bibinfo{person}{Sheelagh Carpendale}.} \bibinfo{year}{2007}\natexlab{}.
\newblock \showarticletitle{Interactive tree comparison for co-located
  collaborative information visualization}.
\newblock \bibinfo{journal}{\emph{IEEE Transactions on Visualization and
  Computer Graphics}} \bibinfo{volume}{13}, \bibinfo{number}{6}
  (\bibinfo{year}{2007}), \bibinfo{pages}{1232--1239}.
\newblock


\bibitem[\protect\citeauthoryear{Isenberg, Elmqvist, Scholtz, Cernea, Ma, and
  Hagen}{Isenberg et~al\mbox{.}}{2011a}]%
        {isenberg2011collaborative}
\bibfield{author}{\bibinfo{person}{Petra Isenberg}, \bibinfo{person}{Niklas
  Elmqvist}, \bibinfo{person}{Jean Scholtz}, \bibinfo{person}{Daniel Cernea},
  \bibinfo{person}{Kwan-Liu Ma}, {and} \bibinfo{person}{Hans Hagen}.}
  \bibinfo{year}{2011}\natexlab{a}.
\newblock \showarticletitle{Collaborative visualization: Definition,
  challenges, and research agenda}.
\newblock \bibinfo{journal}{\emph{Information Visualization}}
  \bibinfo{volume}{10}, \bibinfo{number}{4} (\bibinfo{year}{2011}),
  \bibinfo{pages}{310--326}.
\newblock


\bibitem[\protect\citeauthoryear{Isenberg and Fisher}{Isenberg and
  Fisher}{2009}]%
        {isenberg2009collaborative}
\bibfield{author}{\bibinfo{person}{Petra Isenberg} {and}
  \bibinfo{person}{Danyel Fisher}.} \bibinfo{year}{2009}\natexlab{}.
\newblock \showarticletitle{Collaborative brushing and linking for co-located
  visual analytics of document collections}. In
  \bibinfo{booktitle}{\emph{Computer Graphics Forum}},
  Vol.~\bibinfo{volume}{28}. Wiley Online Library, \bibinfo{pages}{1031--1038}.
\newblock


\bibitem[\protect\citeauthoryear{Isenberg, Fisher, Paul, Morris, Inkpen, and
  Czerwinski}{Isenberg et~al\mbox{.}}{2011b}]%
        {isenberg2011co}
\bibfield{author}{\bibinfo{person}{Petra Isenberg}, \bibinfo{person}{Danyel
  Fisher}, \bibinfo{person}{Sharoda~A Paul}, \bibinfo{person}{Meredith~Ringel
  Morris}, \bibinfo{person}{Kori Inkpen}, {and} \bibinfo{person}{Mary
  Czerwinski}.} \bibinfo{year}{2011}\natexlab{b}.
\newblock \showarticletitle{Co-located collaborative visual analytics around a
  tabletop display}.
\newblock \bibinfo{journal}{\emph{IEEE TVCG}} \bibinfo{volume}{18},
  \bibinfo{number}{5} (\bibinfo{year}{2011}), \bibinfo{pages}{689--702}.
\newblock


\bibitem[\protect\citeauthoryear{{Itakura}}{{Itakura}}{1975}]%
        {itakura}
\bibfield{author}{\bibinfo{person}{F. {Itakura}}.}
  \bibinfo{year}{1975}\natexlab{}.
\newblock \showarticletitle{Minimum prediction residual principle applied to
  speech recognition}.
\newblock \bibinfo{journal}{\emph{IEEE Transactions on Acoustics, Speech, and
  Signal Processing}} \bibinfo{volume}{23}, \bibinfo{number}{1}
  (\bibinfo{year}{1975}), \bibinfo{pages}{67--72}.
\newblock


\bibitem[\protect\citeauthoryear{Javed, McDonnel, and Elmqvist}{Javed
  et~al\mbox{.}}{2010}]%
        {javed2010graphical}
\bibfield{author}{\bibinfo{person}{Waqas Javed}, \bibinfo{person}{Bryan
  McDonnel}, {and} \bibinfo{person}{Niklas Elmqvist}.}
  \bibinfo{year}{2010}\natexlab{}.
\newblock \showarticletitle{Graphical perception of multiple time series}.
\newblock \bibinfo{journal}{\emph{IEEE TVCG}} \bibinfo{volume}{16},
  \bibinfo{number}{6} (\bibinfo{year}{2010}), \bibinfo{pages}{927--934}.
\newblock


\bibitem[\protect\citeauthoryear{Kincaid and Lam}{Kincaid and Lam}{2006}]%
        {kincaid2006line}
\bibfield{author}{\bibinfo{person}{Robert Kincaid} {and} \bibinfo{person}{Heidi
  Lam}.} \bibinfo{year}{2006}\natexlab{}.
\newblock \showarticletitle{Line graph explorer: scalable display of line
  graphs using focus+ context}. In \bibinfo{booktitle}{\emph{Proceedings of the
  working conference on Advanced visual interfaces}}.
  \bibinfo{pages}{404--411}.
\newblock


\bibitem[\protect\citeauthoryear{Li, Monroe, and Jurafsky}{Li
  et~al\mbox{.}}{2016}]%
        {li2016understanding}
\bibfield{author}{\bibinfo{person}{Jiwei Li}, \bibinfo{person}{Will Monroe},
  {and} \bibinfo{person}{Dan Jurafsky}.} \bibinfo{year}{2016}\natexlab{}.
\newblock \showarticletitle{Understanding neural networks through
  representation erasure}.
\newblock \bibinfo{journal}{\emph{arXiv preprint arXiv:1612.08220}}
  (\bibinfo{year}{2016}).
\newblock


\bibitem[\protect\citeauthoryear{Liu, Cui, Jin, Guo, and Qu}{Liu
  et~al\mbox{.}}{2018a}]%
        {liu2018deeptracker}
\bibfield{author}{\bibinfo{person}{Dongyu Liu}, \bibinfo{person}{Weiwei Cui},
  \bibinfo{person}{Kai Jin}, \bibinfo{person}{Yuxiao Guo}, {and}
  \bibinfo{person}{Huamin Qu}.} \bibinfo{year}{2018}\natexlab{a}.
\newblock \showarticletitle{Deeptracker: Visualizing the training process of
  convolutional neural networks}.
\newblock \bibinfo{journal}{\emph{ACM Transactions on Intelligent Systems and
  Technology (TIST)}} \bibinfo{volume}{10}, \bibinfo{number}{1}
  (\bibinfo{year}{2018}), \bibinfo{pages}{1--25}.
\newblock


\bibitem[\protect\citeauthoryear{Liu, Xiao, Browne, Yang, and Dow}{Liu
  et~al\mbox{.}}{2018b}]%
        {liu2018consensus}
\bibfield{author}{\bibinfo{person}{Weichen Liu}, \bibinfo{person}{Sijia Xiao},
  \bibinfo{person}{Jacob~T. Browne}, \bibinfo{person}{Ming Yang}, {and}
  \bibinfo{person}{Steven~P. Dow}.} \bibinfo{year}{2018}\natexlab{b}.
\newblock \showarticletitle{ConsensUs: Supporting Multi-Criteria Group
  Decisions by Visualizing Points of Disagreement}.
\newblock  \bibinfo{volume}{1}, \bibinfo{number}{1} (\bibinfo{year}{2018}).
\newblock
\showISSN{2469-7818}
\urldef\tempurl%
\url{https://doi.org/10.1145/3159649}
\showDOI{\tempurl}


\bibitem[\protect\citeauthoryear{Ludwig, Hilbert, and Pipek}{Ludwig
  et~al\mbox{.}}{2015}]%
        {ludwig2015collaborative}
\bibfield{author}{\bibinfo{person}{Thomas Ludwig}, \bibinfo{person}{Tino
  Hilbert}, {and} \bibinfo{person}{Volkmar Pipek}.}
  \bibinfo{year}{2015}\natexlab{}.
\newblock \showarticletitle{Collaborative visualization for supporting the
  analysis of mobile device data}. In \bibinfo{booktitle}{\emph{2015 ECSCW}}.
  Springer, \bibinfo{pages}{305--316}.
\newblock


\bibitem[\protect\citeauthoryear{Maaten and Hinton}{Maaten and Hinton}{2008}]%
        {maaten2008visualizing}
\bibfield{author}{\bibinfo{person}{Laurens van~der Maaten} {and}
  \bibinfo{person}{Geoffrey Hinton}.} \bibinfo{year}{2008}\natexlab{}.
\newblock \showarticletitle{Visualizing data using t-SNE}.
\newblock \bibinfo{journal}{\emph{Journal of machine learning research}}
  \bibinfo{volume}{9}, \bibinfo{number}{Nov} (\bibinfo{year}{2008}),
  \bibinfo{pages}{2579--2605}.
\newblock


\bibitem[\protect\citeauthoryear{Mahyar and Tory}{Mahyar and Tory}{2014}]%
        {mahyar2014supporting}
\bibfield{author}{\bibinfo{person}{Narges Mahyar} {and}
  \bibinfo{person}{Melanie Tory}.} \bibinfo{year}{2014}\natexlab{}.
\newblock \showarticletitle{Supporting communication and coordination in
  collaborative sensemaking}.
\newblock \bibinfo{journal}{\emph{IEEE TVCG}} \bibinfo{volume}{20},
  \bibinfo{number}{12} (\bibinfo{year}{2014}), \bibinfo{pages}{1633--1642}.
\newblock


\bibitem[\protect\citeauthoryear{McLachlan, Munzner, Koutsofios, and
  North}{McLachlan et~al\mbox{.}}{2008}]%
        {mclachlan2008liverac}
\bibfield{author}{\bibinfo{person}{Peter McLachlan}, \bibinfo{person}{Tamara
  Munzner}, \bibinfo{person}{Eleftherios Koutsofios}, {and}
  \bibinfo{person}{Stephen North}.} \bibinfo{year}{2008}\natexlab{}.
\newblock \showarticletitle{LiveRAC: interactive visual exploration of system
  management time-series data}. In \bibinfo{booktitle}{\emph{Proceedings of the
  SIGCHI Conference on Human Factors in Computing Systems}}.
  \bibinfo{pages}{1483--1492}.
\newblock


\bibitem[\protect\citeauthoryear{Miceli, Schuessler, and Yang}{Miceli
  et~al\mbox{.}}{2020}]%
        {miceli2020between}
\bibfield{author}{\bibinfo{person}{Milagros Miceli}, \bibinfo{person}{Martin
  Schuessler}, {and} \bibinfo{person}{Tianling Yang}.}
  \bibinfo{year}{2020}\natexlab{}.
\newblock \showarticletitle{Between Subjectivity and Imposition: Power Dynamics
  in Data Annotation for Computer Vision}.
\newblock \bibinfo{journal}{\emph{Computer Supported Cooperative Work (CSCW)}}
  \bibinfo{volume}{4}, \bibinfo{number}{2} (\bibinfo{year}{2020}),
  \bibinfo{pages}{1--25}.
\newblock


\bibitem[\protect\citeauthoryear{Morris}{Morris}{2013}]%
        {morris2013collaborative}
\bibfield{author}{\bibinfo{person}{Meredith~Ringel Morris}.}
  \bibinfo{year}{2013}\natexlab{}.
\newblock \showarticletitle{Collaborative search revisited}. In
  \bibinfo{booktitle}{\emph{Proceedings of the 2013 conference on Computer
  supported cooperative work}}. \bibinfo{pages}{1181--1192}.
\newblock


\bibitem[\protect\citeauthoryear{Morris and Horvitz}{Morris and
  Horvitz}{2007}]%
        {morris2007searchtogether}
\bibfield{author}{\bibinfo{person}{Meredith~Ringel Morris} {and}
  \bibinfo{person}{Eric Horvitz}.} \bibinfo{year}{2007}\natexlab{}.
\newblock \showarticletitle{SearchTogether: an interface for collaborative web
  search}. In \bibinfo{booktitle}{\emph{Proceedings of the 20th annual ACM
  symposium on User interface software and technology}}.
  \bibinfo{pages}{3--12}.
\newblock


\bibitem[\protect\citeauthoryear{Pena, de~Assis, and Proen{\c{c}}a}{Pena
  et~al\mbox{.}}{2013}]%
        {pena2013anomaly}
\bibfield{author}{\bibinfo{person}{Eduardo~HM Pena}, \bibinfo{person}{Marcos~VO
  de Assis}, {and} \bibinfo{person}{Mario~Lemes Proen{\c{c}}a}.}
  \bibinfo{year}{2013}\natexlab{}.
\newblock \showarticletitle{Anomaly detection using forecasting methods arima
  and hwds}. In \bibinfo{booktitle}{\emph{2013 IEEE International Conference of
  the Chilean Computer Science Society (SCCC)}}. \bibinfo{pages}{63--66}.
\newblock


\bibitem[\protect\citeauthoryear{Qu, Chan, Xu, Chung, Lau, and Guo}{Qu
  et~al\mbox{.}}{2007}]%
        {qu2007visual}
\bibfield{author}{\bibinfo{person}{Huamin Qu}, \bibinfo{person}{Wing-Yi Chan},
  \bibinfo{person}{Anbang Xu}, \bibinfo{person}{Kai-Lun Chung},
  \bibinfo{person}{Kai-Hon Lau}, {and} \bibinfo{person}{Ping Guo}.}
  \bibinfo{year}{2007}\natexlab{}.
\newblock \showarticletitle{Visual analysis of the air pollution problem in
  Hong Kong}.
\newblock \bibinfo{journal}{\emph{IEEE TVCG}} \bibinfo{volume}{13},
  \bibinfo{number}{6} (\bibinfo{year}{2007}), \bibinfo{pages}{1408--1415}.
\newblock


\bibitem[\protect\citeauthoryear{{Sakoe} and {Chiba}}{{Sakoe} and
  {Chiba}}{1978}]%
        {sakoechiba}
\bibfield{author}{\bibinfo{person}{H. {Sakoe}} {and} \bibinfo{person}{S.
  {Chiba}}.} \bibinfo{year}{1978}\natexlab{}.
\newblock \showarticletitle{Dynamic programming algorithm optimization for
  spoken word recognition}.
\newblock \bibinfo{journal}{\emph{IEEE Transactions on Acoustics, Speech, and
  Signal Processing}} \bibinfo{volume}{26}, \bibinfo{number}{1}
  (\bibinfo{year}{1978}), \bibinfo{pages}{43--49}.
\newblock


\bibitem[\protect\citeauthoryear{Sarkar, Spott, Blackwell, and Jamnik}{Sarkar
  et~al\mbox{.}}{[n.d.]}]%
        {sarkar2016visual}
\bibfield{author}{\bibinfo{person}{Advait Sarkar}, \bibinfo{person}{Martin
  Spott}, \bibinfo{person}{Alan~F Blackwell}, {and} \bibinfo{person}{Mateja
  Jamnik}.} \bibinfo{year}{[n.d.]}\natexlab{}.
\newblock \showarticletitle{Visual discovery and model-driven explanation of
  time series patterns}. In \bibinfo{booktitle}{\emph{2016 IEEE Symposium on
  Visual Languages and Human-Centric Computing}}. \bibinfo{pages}{78--86}.
\newblock


\bibitem[\protect\citeauthoryear{Schlegel, Arnout, El-Assady, Oelke, and
  Keim}{Schlegel et~al\mbox{.}}{2019}]%
        {schlegel2019towards}
\bibfield{author}{\bibinfo{person}{Udo Schlegel}, \bibinfo{person}{Hiba
  Arnout}, \bibinfo{person}{Mennatallah El-Assady}, \bibinfo{person}{Daniela
  Oelke}, {and} \bibinfo{person}{Daniel~A Keim}.}
  \bibinfo{year}{2019}\natexlab{}.
\newblock \showarticletitle{Towards a rigorous evaluation of XAI Methods on
  Time Series}.
\newblock \bibinfo{journal}{\emph{arXiv preprint arXiv:1909.07082}}
  (\bibinfo{year}{2019}).
\newblock


\bibitem[\protect\citeauthoryear{Settles}{Settles}{2011}]%
        {settles2011closing}
\bibfield{author}{\bibinfo{person}{Burr Settles}.}
  \bibinfo{year}{2011}\natexlab{}.
\newblock \showarticletitle{Closing the loop: Fast, interactive semi-supervised
  annotation with queries on features and instances}. In
  \bibinfo{booktitle}{\emph{Proceedings of the 2011 Conference on Empirical
  Methods in Natural Language Processing}}. \bibinfo{pages}{1467--1478}.
\newblock


\bibitem[\protect\citeauthoryear{Shah}{Shah}{2012}]%
        {shah2012collaborative}
\bibfield{author}{\bibinfo{person}{Chirag Shah}.}
  \bibinfo{year}{2012}\natexlab{}.
\newblock \bibinfo{booktitle}{\emph{Collaborative information seeking: The art
  and science of making the whole greater than the sum of all}}.
  Vol.~\bibinfo{volume}{34}.
\newblock \bibinfo{publisher}{Springer Science \& Business Media}.
\newblock


\bibitem[\protect\citeauthoryear{Shneiderman}{Shneiderman}{2003}]%
        {shneiderman2003eyes}
\bibfield{author}{\bibinfo{person}{Ben Shneiderman}.}
  \bibinfo{year}{2003}\natexlab{}.
\newblock \showarticletitle{The eyes have it: A task by data type taxonomy for
  information visualizations}.
\newblock In \bibinfo{booktitle}{\emph{The craft of information
  visualization}}. \bibinfo{publisher}{Elsevier}, \bibinfo{pages}{364--371}.
\newblock


\bibitem[\protect\citeauthoryear{Smith, Sala, Kanter, and Veeramachaneni}{Smith
  et~al\mbox{.}}{2020}]%
        {smith2020machine}
\bibfield{author}{\bibinfo{person}{Micah~J Smith}, \bibinfo{person}{Carles
  Sala}, \bibinfo{person}{James~Max Kanter}, {and} \bibinfo{person}{Kalyan
  Veeramachaneni}.} \bibinfo{year}{2020}\natexlab{}.
\newblock \showarticletitle{The machine learning bazaar: Harnessing the ML
  ecosystem for effective system development}. In
  \bibinfo{booktitle}{\emph{2020 ACM SIGMOD}}. \bibinfo{pages}{785--800}.
\newblock


\bibitem[\protect\citeauthoryear{Torkildson, Starbird, and Aragon}{Torkildson
  et~al\mbox{.}}{2014}]%
        {torkildson2014analysis}
\bibfield{author}{\bibinfo{person}{Megan~K Torkildson}, \bibinfo{person}{Kate
  Starbird}, {and} \bibinfo{person}{Cecilia Aragon}.}
  \bibinfo{year}{2014}\natexlab{}.
\newblock \showarticletitle{Analysis and visualization of sentiment and emotion
  on crisis tweets}. In \bibinfo{booktitle}{\emph{International conference on
  cooperative design, visualization and engineering}}. Springer,
  \bibinfo{pages}{64--67}.
\newblock


\bibitem[\protect\citeauthoryear{Van~Wijk and Van~Selow}{Van~Wijk and
  Van~Selow}{1999}]%
        {van1999cluster}
\bibfield{author}{\bibinfo{person}{Jarke~J Van~Wijk} {and}
  \bibinfo{person}{Edward~R Van~Selow}.} \bibinfo{year}{1999}\natexlab{}.
\newblock \showarticletitle{Cluster and calendar based visualization of time
  series data}. In \bibinfo{booktitle}{\emph{1999 InfoVis}}.
  \bibinfo{pages}{4--9}.
\newblock


\bibitem[\protect\citeauthoryear{Viegas and Wattenberg}{Viegas and
  Wattenberg}{2006}]%
        {viegas2006communication}
\bibfield{author}{\bibinfo{person}{Fernanda~B Viegas} {and}
  \bibinfo{person}{Martin Wattenberg}.} \bibinfo{year}{2006}\natexlab{}.
\newblock \showarticletitle{Communication-minded visualization: A call to
  action}.
\newblock \bibinfo{journal}{\emph{IBM Systems Journal}} \bibinfo{volume}{45},
  \bibinfo{number}{4} (\bibinfo{year}{2006}), \bibinfo{pages}{801}.
\newblock


\bibitem[\protect\citeauthoryear{Vijaymeena and Kavitha}{Vijaymeena and
  Kavitha}{2016}]%
        {vijaymeena2016survey}
\bibfield{author}{\bibinfo{person}{MK Vijaymeena} {and} \bibinfo{person}{K
  Kavitha}.} \bibinfo{year}{2016}\natexlab{}.
\newblock \showarticletitle{A survey on similarity measures in text mining}.
\newblock \bibinfo{journal}{\emph{Machine Learning and Applications: An
  International Journal}} \bibinfo{volume}{3}, \bibinfo{number}{2}
  (\bibinfo{year}{2016}), \bibinfo{pages}{19--28}.
\newblock


\bibitem[\protect\citeauthoryear{Wang, Yang, Abdul, and Lim}{Wang
  et~al\mbox{.}}{2019}]%
        {wang2019designing}
\bibfield{author}{\bibinfo{person}{Danding Wang}, \bibinfo{person}{Qian Yang},
  \bibinfo{person}{Ashraf Abdul}, {and} \bibinfo{person}{Brian~Y Lim}.}
  \bibinfo{year}{2019}\natexlab{}.
\newblock \showarticletitle{Designing theory-driven user-centric explainable
  AI}. In \bibinfo{booktitle}{\emph{Proceedings of the SIGCHI Conference on
  Human Factors in Computing Systems}}. \bibinfo{pages}{1--15}.
\newblock


\bibitem[\protect\citeauthoryear{Ware}{Ware}{2019}]%
        {ware2019information}
\bibfield{author}{\bibinfo{person}{Colin Ware}.}
  \bibinfo{year}{2019}\natexlab{}.
\newblock \bibinfo{booktitle}{\emph{Information visualization: perception for
  design}}.
\newblock \bibinfo{publisher}{Morgan Kaufmann}.
\newblock


\bibitem[\protect\citeauthoryear{Weber, Alexa, and M{\"u}ller}{Weber
  et~al\mbox{.}}{2001}]%
        {weber2001visualizing}
\bibfield{author}{\bibinfo{person}{Marc Weber}, \bibinfo{person}{Marc Alexa},
  {and} \bibinfo{person}{Wolfgang M{\"u}ller}.}
  \bibinfo{year}{2001}\natexlab{}.
\newblock \showarticletitle{Visualizing time-series on spirals}. In
  \bibinfo{booktitle}{\emph{2001 InfoVis}}, Vol.~\bibinfo{volume}{1}.
  \bibinfo{pages}{7--14}.
\newblock


\bibitem[\protect\citeauthoryear{Willett, Heer, Hellerstein, and
  Agrawala}{Willett et~al\mbox{.}}{2011}]%
        {willett2011commentspace}
\bibfield{author}{\bibinfo{person}{Wesley Willett}, \bibinfo{person}{Jeffrey
  Heer}, \bibinfo{person}{Joseph Hellerstein}, {and} \bibinfo{person}{Maneesh
  Agrawala}.} \bibinfo{year}{2011}\natexlab{}.
\newblock \showarticletitle{CommentSpace: structured support for collaborative
  visual analysis}. In \bibinfo{booktitle}{\emph{Proceedings of the SIGCHI
  conference on Human Factors in Computing Systems}}.
  \bibinfo{pages}{3131--3140}.
\newblock


\bibitem[\protect\citeauthoryear{Xia, Zhao, Sheeley, Christopher, Wang, Guo,
  Zhang, Ebert, Chen, and Qian}{Xia et~al\mbox{.}}{2014}]%
        {xia2014annotatedtimetree}
\bibfield{author}{\bibinfo{person}{Jing Xia}, \bibinfo{person}{Jieqiong Zhao},
  \bibinfo{person}{Isaac Sheeley}, \bibinfo{person}{Joseph Christopher},
  \bibinfo{person}{Qiaoying Wang}, \bibinfo{person}{Chen Guo},
  \bibinfo{person}{Jiawei Zhang}, \bibinfo{person}{David~S Ebert},
  \bibinfo{person}{Yingjie~Victor Chen}, {and} \bibinfo{person}{Zhenyu~Cheryl
  Qian}.} \bibinfo{year}{2014}\natexlab{}.
\newblock \showarticletitle{AnnotatedTimeTree: Visualization and annotation of
  news text and other heterogeneous document collections}. In
  \bibinfo{booktitle}{\emph{2014 IEEE VAST}}. \bibinfo{pages}{337--338}.
\newblock


\bibitem[\protect\citeauthoryear{Xie, Xu, and Mueller}{Xie
  et~al\mbox{.}}{2018}]%
        {xie2018visual}
\bibfield{author}{\bibinfo{person}{Cong Xie}, \bibinfo{person}{Wei Xu}, {and}
  \bibinfo{person}{Klaus Mueller}.} \bibinfo{year}{2018}\natexlab{}.
\newblock \showarticletitle{A visual analytics framework for the detection of
  anomalous call stack trees in high performance computing applications}.
\newblock \bibinfo{journal}{\emph{IEEE TVCG}} \bibinfo{volume}{25},
  \bibinfo{number}{1} (\bibinfo{year}{2018}), \bibinfo{pages}{215--224}.
\newblock


\bibitem[\protect\citeauthoryear{Xu, Guo, Cao, Gotz, Xu, Qu, Yao, and Chen}{Xu
  et~al\mbox{.}}{2018}]%
        {xu2018ecglens}
\bibfield{author}{\bibinfo{person}{Ke Xu}, \bibinfo{person}{Shunan Guo},
  \bibinfo{person}{Nan Cao}, \bibinfo{person}{David Gotz},
  \bibinfo{person}{Aiwen Xu}, \bibinfo{person}{Huamin Qu},
  \bibinfo{person}{Zhenjie Yao}, {and} \bibinfo{person}{Yixin Chen}.}
  \bibinfo{year}{2018}\natexlab{}.
\newblock \showarticletitle{Ecglens: Interactive visual exploration of large
  scale ecg data for arrhythmia detection}. In
  \bibinfo{booktitle}{\emph{Proceedings of the SIGCHI Conference on Human
  Factors in Computing Systems}}. \bibinfo{pages}{1--12}.
\newblock


\bibitem[\protect\citeauthoryear{Xu, Wang, Yang, Wang, Qiao, Qin, Xu, Zhang,
  and Qu}{Xu et~al\mbox{.}}{2019}]%
        {xu2019clouddet}
\bibfield{author}{\bibinfo{person}{Ke Xu}, \bibinfo{person}{Yun Wang},
  \bibinfo{person}{Leni Yang}, \bibinfo{person}{Yifang Wang},
  \bibinfo{person}{Bo Qiao}, \bibinfo{person}{Si Qin}, \bibinfo{person}{Yong
  Xu}, \bibinfo{person}{Haidong Zhang}, {and} \bibinfo{person}{Huamin Qu}.}
  \bibinfo{year}{2019}\natexlab{}.
\newblock \showarticletitle{CloudDet: Interactive Visual Analysis of Anomalous
  Performances in Cloud Computing Systems}.
\newblock \bibinfo{journal}{\emph{IEEE TVCG}} \bibinfo{volume}{26},
  \bibinfo{number}{1} (\bibinfo{year}{2019}), \bibinfo{pages}{1107--1117}.
\newblock


\bibitem[\protect\citeauthoryear{Zhao, Glueck, Isenberg, Chevalier, and
  Khan}{Zhao et~al\mbox{.}}{2017}]%
        {zhao2017supporting}
\bibfield{author}{\bibinfo{person}{Jian Zhao}, \bibinfo{person}{Michael
  Glueck}, \bibinfo{person}{Petra Isenberg}, \bibinfo{person}{Fanny Chevalier},
  {and} \bibinfo{person}{Azam Khan}.} \bibinfo{year}{2017}\natexlab{}.
\newblock \showarticletitle{Supporting handoff in asynchronous collaborative
  sensemaking using knowledge-transfer graphs}.
\newblock \bibinfo{journal}{\emph{IEEE TVCG}} \bibinfo{volume}{24},
  \bibinfo{number}{1} (\bibinfo{year}{2017}), \bibinfo{pages}{340--350}.
\newblock


\bibitem[\protect\citeauthoryear{Zheng, Li, and Zhao}{Zheng
  et~al\mbox{.}}{2016}]%
        {zheng2016self}
\bibfield{author}{\bibinfo{person}{Dequan Zheng}, \bibinfo{person}{Fenghuan
  Li}, {and} \bibinfo{person}{Tiejun Zhao}.} \bibinfo{year}{2016}\natexlab{}.
\newblock \showarticletitle{Self-adaptive statistical process control for
  anomaly detection in time series}.
\newblock \bibinfo{journal}{\emph{Expert Systems with Applications}}
  \bibinfo{volume}{57} (\bibinfo{year}{2016}), \bibinfo{pages}{324--336}.
\newblock


\bibitem[\protect\citeauthoryear{Zhou, Liu, Hooi, Cheng, and Ye}{Zhou
  et~al\mbox{.}}{2019}]%
        {zhou2019beatgan}
\bibfield{author}{\bibinfo{person}{Bin Zhou}, \bibinfo{person}{Shenghua Liu},
  \bibinfo{person}{Bryan Hooi}, \bibinfo{person}{Xueqi Cheng}, {and}
  \bibinfo{person}{Jing Ye}.} \bibinfo{year}{2019}\natexlab{}.
\newblock \showarticletitle{BeatGAN: Anomalous Rhythm Detection using
  Adversarially Generated Time Series}. In \bibinfo{booktitle}{\emph{IJCAI}}.
  \bibinfo{pages}{4433--4439}.
\newblock


\bibitem[\protect\citeauthoryear{Zytek, Liu, Vaithianathan, and
  Veeramachaneni}{Zytek et~al\mbox{.}}{2021}]%
        {zytek2021sibyl}
\bibfield{author}{\bibinfo{person}{Alexandra Zytek}, \bibinfo{person}{Dongyu
  Liu}, \bibinfo{person}{Rhema Vaithianathan}, {and} \bibinfo{person}{Kalyan
  Veeramachaneni}.} \bibinfo{year}{2021}\natexlab{}.
\newblock \showarticletitle{Sibyl: Understanding and Addressing the Usability
  Challenges of Machine Learning In High-Stakes Decision Making}.
\newblock \bibinfo{journal}{\emph{IEEE Transactions on Visualization and
  Computer Graphics}} (\bibinfo{year}{2021}).
\newblock


\end{thebibliography}

\end{document}